\documentclass{WileyMSP-template}

\usepackage{bm,color,amsmath,txfonts}
\usepackage{graphicx}
\usepackage{siunitx}

\usepackage{verbatim}
\usepackage{dcolumn}
\usepackage{bm}
\usepackage{epsf}
\usepackage{xcolor}
\usepackage{hyperref}
\usepackage{hhline}
\usepackage{float}
\usepackage{enumerate}
\usepackage{bbm}
\usepackage{lipsum}
\usepackage{cite}
\usepackage{graphicx}
\usepackage{subcaption}
\usepackage{tikz}

\usepackage{mathrsfs}

\begin{document}

\pagestyle{fancy}
\rhead{\includegraphics[width=2.5cm]{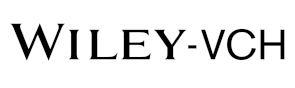}}


\title{Integrated Ring-based Quantum Key Distribution with Weak Measurement Enhanced Fiber-Optic Sensing Disturbance Magnitude and Location}


\maketitle


\author{Weiqian Zhao}
\author{Wenzhao Huang}
\author{Zifu Su}
\author{Fangyuan Li}
\author{Qirong Jiang}
\author{Cheng Yuan}
\author{Yafei Yu}
\author{Jindong Wang}
	
\dedication{}

\begin{affiliations}
Weiqian Zhao, Wenzhao Huang, Zifu Su, Fangyuan Li, Qirong Jiang, Cheng Yuan\\
Guangdong Provincial Key Laboratory of Nanophotonic Functional Materials and Devices, School of Optoelectronic Science and Engineering,  South China Normal University, Guangzhou 510006, China\\

Yafei Yu\\
Guangdong Provincial Key Laboratory of Nanophotonic Functional Materials and Devices, School of Optoelectronic Science and Engineering,  South China Normal University, Guangzhou 510006, China\\
Quantum Science Center of Guangdong-Hong Kong-Macao Greater Bay Area, Shenzhen-Hong Kong International Science and Technology Park,  NO.3 Binglang Road, Futian District, Shenzhen, Guangdong\\
Email: yuyafei@m.scnu.edu.cn

Jindong Wang\\
Guangdong Provincial Key Laboratory of Quantum Engineering and Quantum Materials, School of Optoelectronic Science and Engineering, South China Normal University, Guangzhou 510006, China\\
 Guangdong Basic Research Center of Excellence for Structure and Fundamental Interactions of Matter, 
School of Optoelectronic Science and Engineering, 
South China Normal University, Guangzhou 510006, China\\
Email: wangjindong@m.scnu.edu.cn
\end{affiliations}

	
\begin{abstract}

The deep integration of quantum communication and fiber-optic sensing is pivotal for the development of next-generation multifunctional and highly reliable secure information infrastructure. 
Here, we present a Sagnac-loop integrated system (SLIS) that, for the first time, combines ring-based quantum key distribution (QKD) with fiber-based weak measurement (WM) enhanced sensing and disturbance localization capabilities. In the event of communication interruption due to external disturbances, the SLIS seamlessly switches to perception system, employing interference measurement and WM techniques to monitor channel disturbances. By integrating null-frequencies localization (NFL) mode, the system precisely determines the disturbance location, enabling rapid identification of security vulnerabilities along the link.
Experimental results demonstrate that, over a 30 km Sagnac loop channel, the SLIS achieves a raw key generation rate of 22.4 kbps with stable operation and clear scalability toward network expansion. In terms of perception performance, the SLIS exhibits strong capability for both dynamic and quasi-static disturbances. For dynamic perturbations, the system detects transient impacts and PZT-driven frequency variations down to 100 Hz, and enables long-distance localization via NFL alignment, with improved localization performance as the disturbance position moves farther away along the loop. For quasi-static disturbances, gravitational changes as small as 100 g are resolved, corresponding to a time-delay variation of 9.81 as.
This work provides a novel technical pathway toward self-diagnosing, robust quantum networks through integrated communication and sensing functionalities.

\end{abstract}

	
\section{Introduction}
\justifying
	In recent years, with the rapid advancement of quantum information science, quantum communication (QC) has emerged as one of the most promising technologies for practical implementation, attracting extensive attention from researchers worldwide. In particular, the continuous progress of quantum key distribution (QKD) has driven its transition from laboratory demonstrations to large-scale network deployment, establishing it as a cornerstone for future information security infrastructures \cite{QKD_Conjugate_Conding, QKD_BB, QKD_LKG, QKD_PCR, QKD_Advance}. Meanwhile, the growing convergence between quantum and classical communication systems has accelerated the evolution of quantum network (QN) toward higher integration, robustness, and multifunctionality \cite{Qua_Cla_enhanced, Qua_Cla_Wireless, Qua_Cla_Novel, Qua_Cla_Hybrid, Qua_Cla_multicore_fibers, Qua_Cla_interplay, Qua_Cla_network, Qua_Cla_advance_intergrated}. As the scale of QN expands and environmental conditions become increasingly complex, achieving high-sensitivity detection and precise localization of channel disturbances has become a critical challenge for maintaining stable system operation. Against this backdrop, novel system architectures that integrate quantum sensing (QS) theory with fiber-optic sensing technologies have attracted considerable attention \cite{Integrated_Polan, Integrated_Shapiro, Integrated_GuoGC, Integrated_ZhengGH, Integrated_LiYM, Integrated_IoT, Integrated_PanJW}, offering a promising pathway to enhance both the stability and practicality of QC systems.

QKD has achieved remarkable progress in both theoretical security analysis and system implementation, and has been widely deployed across various fiber-optic network environments. In particular, diverse network architectures—such as point-to-point \cite{QKD_NW_P_P_Field, QKD_NW_P_P_Tokyo, QKD_NW_P_P_Swiss, QKD_NW_P_P_upstream }, star \cite{ QKD_NW_Star_access, QKD_NW_Star_LYM, QKD_NW_Star_S_G, QKD_NW_Star_ZGH}, and ring topologies \cite{QKD_NW_Ring_muti, QKD_NW_Ring_study, QKD_NW_Ring_distribution}—have been developed and matured, providing flexible and secure key distribution solutions for different application scenarios. However, as QKD systems move toward large-scale practical deployment, challenges such as environmental disturbances, device aging, and link instability have increasingly become key factors limiting key generation efficiency and system stability \cite{QKD_robust_90km, QKD_robust_Waters, QKD_robust_CV,  QKD_robust_Practical}. Conventional QKD systems typically rely on feedback control or redundant encoding schemes to mitigate such disturbances \cite{QKD_compen_Network, QKD_compen_Simple, QKD_compen_PC, QKD_compen_gigahertz, QKD_compen_PCF}. However, these methods often suffer from delayed response and limited capability in identifying or localizing the disturbance sources.

To overcome this problem, advances in fiber-optic sensing technologies have opened new avenues for achieving real-time, high-sensitivity monitoring of channel disturbances \cite{SenCommu_integrated_overview, SenCommu_integrated_highspeed, SenCommu_integrated_Distributed, SenCommu_integrated_Light}. The deep integration of environmentally sensitive fiber-optic sensing mechanisms with QKD systems offers a promising approach to enable rapid detection and localization of disturbances. Moreover, it can provide critical data support for adaptive compensation and security enhancement, further advancing the implementation of communication–sensing integrated architectures within QN. Additionally, this provides valuable guidance and practical experience for the integration of QC and QS, where unified architectures promise significant advantages in resource sharing, operational stability, and communication security.

Against this background, the concept of integrating sensing techniques with QKD systems was first proposed by Polish researchers as early as 2013 \cite{Integrated_Polan}. Subsequently, Zhuang et al. \cite{Integrated_Shapiro} developed a theoretical framework for distributed QS based on continuous-variable (CV) entanglement, demonstrating its potential for calibrating CV-QKD networks. Building upon the same idea, Xu et al. \cite{Integrated_ZhengGH} and Liu et al. \cite{Integrated_LiYM} experimentally realized the fusion of QS and CV-QKD networks through schemes based on spectral phase monitoring and fiber vibration sensing, respectively, thereby achieving both vibration detection and networked QKD operation. Furthermore, Yin et al. \cite{Integrated_GuoGC} combined an entanglement-based QKD protocol with an entanglement-assisted QS system, enabling effective monitoring of both environmental parameters and potential eavesdropping activities. Chen et al. \cite{Integrated_PanJW} also integrated communication and sensing functions by introducing a phase-compensation frequency calibration link into a twin-field QKD system, thus further advancing the practical realization of communication–sensing convergence.

Despite significant progress in integrated sensing and communication within QKD systems, achieving high-sensitivity detection and precise localization of disturbances across realistic network topologies remains a challenge. Here, we propose a Sagnac-loop integrated system (SLIS) that addresses this issue by introducing a QKD architecture based on the Sagnac-loop, which synergistically integrates weak measurement (WM) and weak value amplification (WVA) \cite{Sagnc_QKD_TD, Sagnc_QKD_Telecom, Sagnc_QKD_LKG, Sagnc_QKD_TF, WM_our_work }. This unique combination leverages the inherent stability of the Sagnac interferometer for robust key distribution, while the WVA mechanism enables exceptional sensitivity to subwavelength disturbances. Moreover, a null-frequencies localization (NFL) method based on interference characteristics allows precise estimation of disturbance positions \cite{ZF_ref}. The proposed architecture thus enables simultaneous secure communication and high-fidelity channel diagnostics, advancing the development of self-aware QN.

Experimental results show that over a 30 km fiber Sagnac-loop, SLIS achieves a raw key generation rate of 22.4 \text{kbps} while maintaining a quantum bit error rate (QBER) below 5\text{\%} under 20 minutes long-term continuous operation, without employing any active compensation. For dynamic disturbance of periodic vibration and transient impact , SLIS can identify vibration frequencies as low as 100 Hz in the 1 Hz–10 kHz range. The localization resolution ranges from 0.02 m at the proximal end to 542.2 m at the distal end of the loop, exhibiting a position-dependent characteristic along the fiber, and and a localization error on the order of hundreds of meters. For quasi-static pressure-induced disturbances, the system detects time-delay variations as small as 9.81 as, corresponding to gravitational changes of approximately 100 g . Compared with conventional approaches, this scheme offers distinct advantages in sensing precision, stability, and system-level integration, while exhibiting strong potential for scalable communication–sensing co-design in future robust QN.

The remainder of this paper is organized as follows. Section 2 introduces the theoretical mechanisms underlying the three operational modules of the SLIS, together with the fundamental principles of their integration. Section 3 presents the experimental results demonstrating the communication, sensing, and localization functionalities of the SLIS. Section 4 concludes the paper by summarizing the proposed SLIS framework.

\section{Theoretical Framework}
\justifying
The proposed SLIS consists of three functional modules: a phase-encoded BB84 QKD module \cite{QKD_BB}, a WM-based module for high-precision fiber-optic time delay estimation under large inherent time delays \cite{WM_our_work}, and a disturbance localization module based on the NFL method \cite{ZF_ref}. Together, the sensing and localization modules enable the seamless integration of communication system with a perception system.

\begin{figure}[htbp]
	\centering
	\includegraphics[width=1\textwidth]{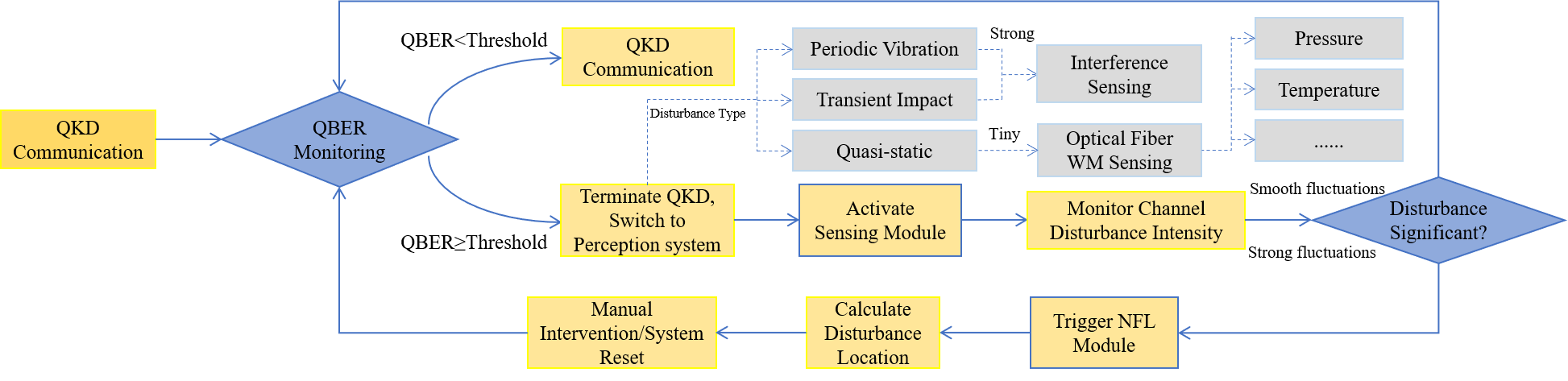}
	\caption{Schematic diagram of the working process of the SLIS.}
	\label{WP}
\end{figure}

The overall working principle and module transition process are illustrated in Figure~\ref{WP}. During normal operation, the SLIS performs QKD while continuously monitoring the QBER. As long as the QBER remains below the predefined security threshold, secure key distribution proceeds without interruption. However, once the QBER exceeds this threshold, the QKD process is automatically halted, and the system switches to the perception system. In this system, the interference and WM-based sensing module is activated to monitor the intensity of various types of channel disturbances. If the disturbance is determined to be significant, the NFL module is triggered to calculate the disturbance position. The resulting localization information then guides either manual intervention or system reset to restore stable QKD operation. This closed-loop workflow allows the system to seamlessly alternate between QC and QS functions, thereby enhancing both the robustness and environmental awareness of the integrated QN.

Sagnac interferometers, owing to their intrinsic capabilities of passive polarization self-alignment and birefringence self-compensation, have become a central architecture in both QKD\cite{ Sagnac_appl_PM, Sagnac_appl_source, Sagnac_appl_CVmodulator, Sagnac_appl_CVfree} and fiber-optic sensing \cite{SenCommu_integrated_Review}. By ensuring that the clockwise (CW) and counterclockwise (CCW) pulses propagate along an identical physical path, the Sagnac configuration naturally suppresses quasi-static phase drifts caused by fiber stress, temperature fluctuations, and long-term environmental disturbances. This inherent stability eliminates the need for active phase control and enables long-term interference coherence, making Sagnac-loop systems particularly attractive for practical fiber-based QC.

\begin{figure}[htbp]
	\centering
	\includegraphics[width=0.6\textwidth]{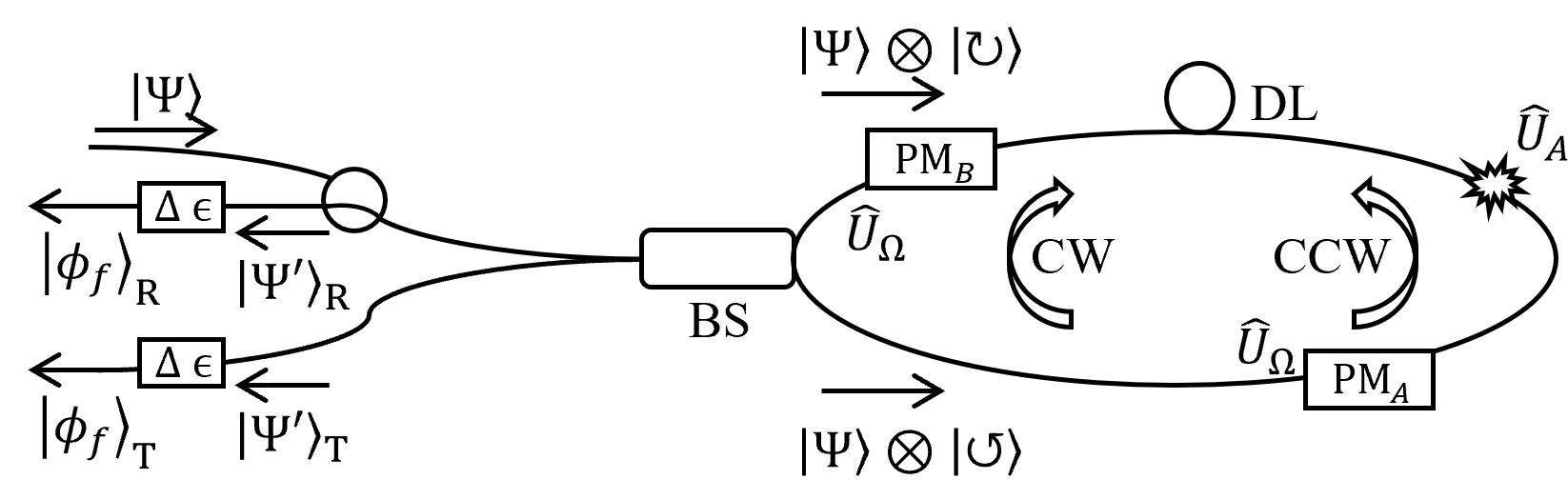}
	\caption{Schematic illustration of the evolution of the optical field within the SLIS.}
	\label{Sch_SLIS}
\end{figure}

Leveraging these advantages, we design the SLIS architecture, as illustrated in Figure~\ref{Sch_SLIS}. Using only a narrow-linewidth laser source, a 50:50 beam splitter (BS), phase modulators (PM), a variable attenuator (ATT), a fiber loop, and a polarization-analysis module, the proposed platform can realize the three categories of functionalities shown in Figure~\ref{WP} within a single Sagnac infrastructure. In the QKD module, a BB84-based time-division phase-encoding scheme is adopted, where CW and CCW pulses are independently modulated by the communicating parties. The communication information is encoded in the global phase difference (GPD) $\Delta\delta$ between the counter-propagating pulses, allowing the system to maintain a compact footprint while ensuring high robustness against environmental disturbances.

\begin{figure}[htbp]
	\centering
	\includegraphics[width=0.6\textwidth]{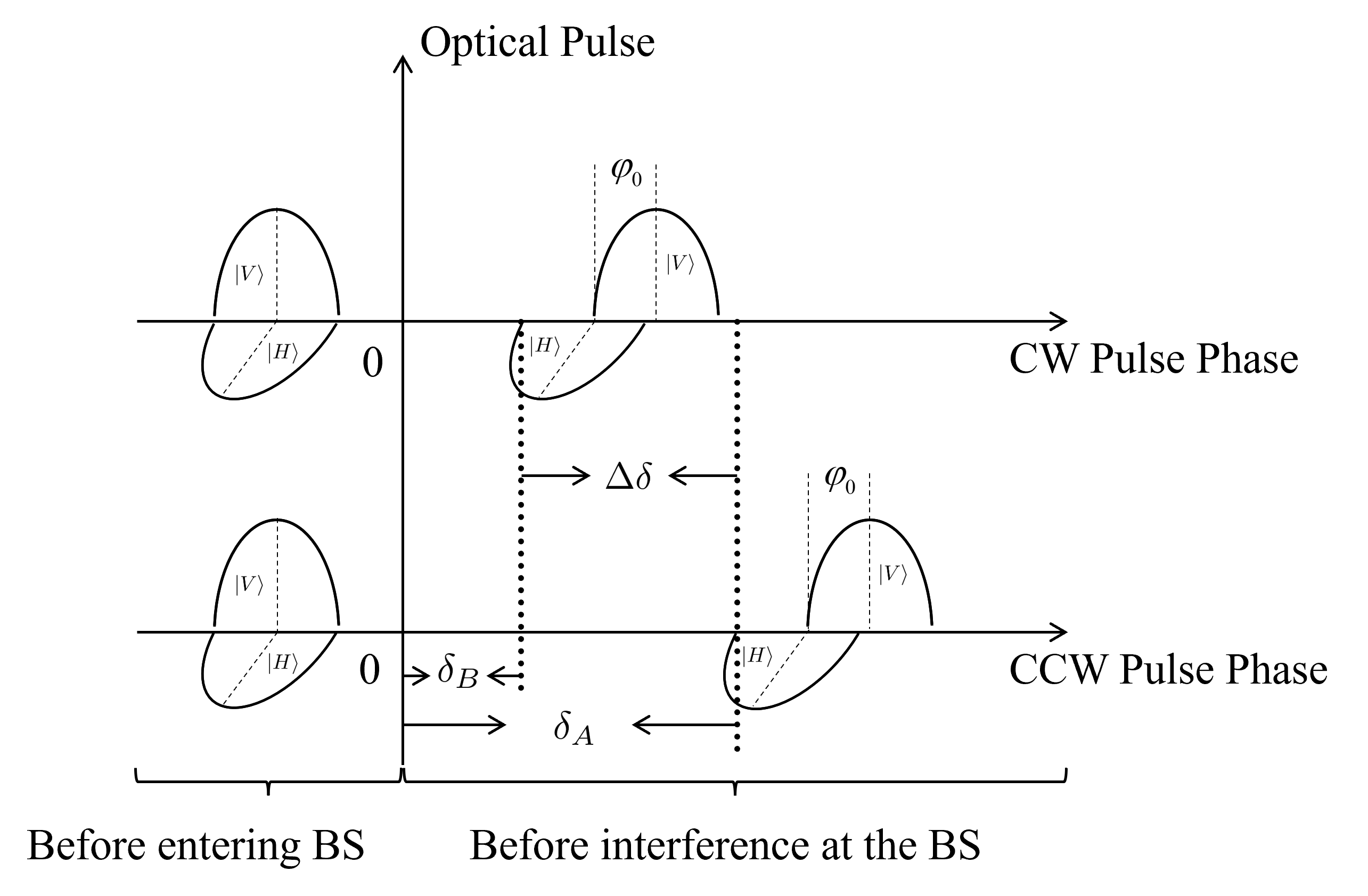}
	\caption{Schematic of the relationship between the communication and sensing carriers in the SLIS.}
	\label{Carriers}
\end{figure}

For the sensing functionality of the SLIS, dynamic disturbances are directly mapped onto variations of the GPD within the loop, allowing time-varying disturbances to be extracted via the same interferometric mechanism employed for QKD. In contrast, quasi-static disturbances are carried in the relative phase difference (RPD) between the two orthogonal polarization components of a single pulse, given by $\varphi_0 = \omega_0 \tau_0$, where $\omega_0$ is the central optical frequency and $\tau_0$ denotes the differential group delay induced by polarization-mode dispersion (PMD), which serves as an intrinsic temporal delay. By introducing a weak interaction form disturbance between the polarization components and applying appropriate post-selection, SLIS attains enhanced sensitivity to external quasi-static disturbances, enabling WM-enhanced fiber sensing. The separation of communication encoding and WM-based sensing channels is schematically illustrated in Figure~\ref{Carriers}.

The origin denotes the time when the optical pulse enters the BS. The negative and positive halves of the time axis represent the initial and returned pulses, respectively. In the absence of external disturbance, the propagating CW and CCW pulses accumulate identical global phases from the common Sagnac-loop path, which cancel upon interference and are thus not shown. They additionally receive distinct global phase modulations $\delta_A$ and $\delta_B$ from the communicating parties,  which satisfy $\Delta\delta = \delta_A - \delta_B$, and acquire a RPD $\varphi_0$ between orthogonal polarizations due to intrinsic birefringence of the fiber channel. The pulses then recombine and interfere at the BS, each comprising both horizontal $|H\rangle$ and vertical $|V\rangle$ polarization components.

In the theoretical analysis of SLIS, the PMs of the two communicating parties independently modulate the global phases of the CW and CCW pulses by $\delta_{A}$ and $\delta_{B}$, respectively. External disturbances introduce an additional phase shift $\Delta\varphi = \omega_0(\tau_0 + \Delta\tau)$, where $\Delta\tau$ denotes the time-delay displacement induced by the disturbance. For quasi-static disturbances, the quantum theoretical description of SLIS is given as follows:

In this process, the polarization degree of freedom of a single optical pulse acts as the interested system. The pre-selection prepares the initial polarization state as $|i\rangle = (|H\rangle + |V\rangle)/\sqrt{2}$. This corresponds to setting the initial polarization of the optical pulse at an angle of $\pi/4$ with respect to effective birefringence axis of the channel, a choice whose rationale is detailed in the Jones-matrix derivation provided in the Supporting Information S1.

The probe system utilizes the frequency distribution degree of freedom of the light beam, with an initial state $|\phi\rangle = \int d\omega f(\omega)|\omega\rangle$, where $\omega$ is the optical frequency and $f(\omega) = (2\pi\sigma^2)^{-1/4}\exp[-(\omega - \omega_0)^2 / (2\sigma^2)]$ is the spectral probability distribution, with $\sigma^2 = \langle \omega^2 \rangle - \langle \omega \rangle^2$ defines the spectral bandwidth of the light. Consequently, the joint initial state of the prepared optical field is:
\begin{equation}
\begin{aligned}
|\Psi \rangle =|i\rangle \otimes |\phi \rangle.
\end{aligned}
\end{equation}

At the BS of the Sagnac-loop, the incident light is divided into two counter-propagating beams: the CW and CCW components. The evolution of these fields through interaction with the external disturbance in the loop is given by the evolution operator:
\begin{equation}
\begin{aligned}
\hat{U}_\text{A} =e^{-i\tau \omega \hat{A}},
\end{aligned}
\end{equation}
here, $\tau = \tau_0+\Delta \tau$ denotes the time delay between the $|H\rangle$ and $|V\rangle$ polarization components within the loop, and $\hat{A} = |H\rangle \langle H| - |V\rangle \langle V|$ represents the observable of the system.

The beam splitting within the Sagnac-loop can be regarded as a pre-selection of the propagation path, with the initial path state given by $|\Omega_i \rangle = (|\circlearrowright\rangle + i|\circlearrowleft\rangle)/\sqrt{2}$, where $|\circlearrowright\rangle$ and $|\circlearrowleft\rangle$ denote the CCW and CW propagation states, respectively. In the Sagnac-loop, the primary influence on the CW and CCW beams arises from the GPD $\Delta \delta$ applied by the communicating parties. Thus, the path evolution is given by:
\begin{equation}
\begin{aligned}
\hat{U}_{\Omega} =e^{-i\Delta \delta \cdot \hat{S}},
\end{aligned}
\end{equation}
where $\hat{S} = |\circlearrowleft\rangle \langle \circlearrowleft| - |\circlearrowright\rangle \langle \circlearrowright|$ represents the observable of the path degree of freedom.

Consequently, the joint state of the beam's polarization and path degrees of freedom after evolution in the Sagnac-loop can be expressed as:
\begin{equation}
\begin{aligned}
|\Psi^{\prime} \rangle 
&= \left( \hat{U}_{A} |i\rangle \right) \otimes \left( \hat{U}_{\Omega} |\Omega_{i} \rangle \right) \otimes |\phi \rangle \\
&= \frac{1}{\sqrt{2}} \left( e^{i\delta_A}|\circlearrowright\rangle + ie^{i \delta_B}|\circlearrowleft\rangle   \right) \otimes \left( \hat{U}_{A} |i\rangle \right) \otimes |\phi \rangle.
\end{aligned}
\end{equation}

After the evolution within the loop, the CW and CCW beams interfere at the BS, a process that can be regarded as a post-selection of the path. The reflected output state is given by
$|\Omega_f\rangle_\text{R} = \left( |\circlearrowright\rangle + i|\circlearrowleft\rangle \right) /\sqrt{2}$,
while the transmitted output state is
$|\Omega_f\rangle_\text{T} = \left( i|\circlearrowright\rangle + |\circlearrowleft\rangle \right) /\sqrt{2}$.
Following the interference at the BS after returning through the Sagnac-loop, the quantum states at the reflected and transmitted ports are:
\begin{subequations}
\begin{equation}
|\Psi^{\prime} \rangle_\text{R} =\frac{e^{i\delta_{A}}+e^{i\delta_{B}}}{2} \hat{U}_\text{A} |i\rangle \otimes |\phi \rangle,
\end{equation}
\begin{equation}
|\Psi^{\prime} \rangle_\text{T} =-i\frac{e^{i\delta_{A}}-e^{i\delta_{B}}}{2} \hat{U}_\text{A} |i\rangle \otimes |\phi \rangle.
\end{equation}
\end{subequations}

Following the path post-selection, the beam undergoes a polarization post-selection, with the post-selected state given by $|f\rangle_\text{T,R} = \left( e^{i\epsilon}|H\rangle -e^{-i\epsilon}|V\rangle \right)/\sqrt{2}$. The post-selected polarization state is nearly orthogonal to the initial preselected state to extract the measurement information, where $\epsilon$ denotes the post-selection angle. After performing the post-selection on the polarization degree of freedom, the final probe state of the beam can be expressed as:
\begin{subequations}
\begin{equation}
|\phi_{f} \rangle_\text{R} = _\text{R}\langle f|\Psi^{\prime} \rangle_\text{R} =\frac{e^{i\delta_{A}}+e^{i\delta_{B}}}{2} \int d\omega f\left( \omega \right) \cos \left( \Delta \varphi -\epsilon \right) |\omega \rangle,
\end{equation}
\begin{equation}
|\phi_{f} \rangle_\text{T} = _\text{T}\langle f|\Psi^{\prime} \rangle_\text{T} =-i\frac{e^{i\delta_{A}}-e^{i\delta_{B}}}{2} \int d\omega f\left( \omega \right) \cos \left( \Delta \varphi -\epsilon \right) |\omega \rangle.
\end{equation}
\end{subequations}

Consequently, when the loop is subjected to external disturbances, the post-selection success probabilities at the reflected and transmitted ports are given by:
\begin{subequations}
\begin{equation}
P_\text{R}=_\text{R}\langle \phi_{f} |\phi_{f} \rangle_\text{R} =\frac{1}{4} \left( 1+\cos \Delta \delta \right) \left\{ 1-e^{-\sigma^{2} \tau^{2}}\cos \left[ 2\left( \Delta \varphi -\epsilon \right) \right] \right\},
\end{equation}
\begin{equation}
P_\text{T}=_\text{T}\langle \phi_{f} |\phi_{f} \rangle_\text{T} =\frac{1}{4} \left( 1-\cos \Delta \delta \right) \left\{ 1-e^{-\sigma^{2} \tau^{2}}\cos \left[ 2\left( \Delta \varphi -\epsilon \right) \right] \right\}.
\end{equation}
\end{subequations}

Specifically, when the communicating parties set the GPD to $\Delta \delta = 0$ or $\pi$, the reflected or transmitted port, respectively, preserves the target RPD $\Delta \varphi$ for measurement. It should be noted that WM sensing remains feasible for the non-key generating GPD settings $\Delta \delta = \pi/2$ and $3\pi/2$, where the optical intensity is evenly distributed between the two output ports. Although the signal-to-noise ratio in this case is lower than that obtained from a single port, this operating regime provides a viable approach for the synchronous extraction of sensing information using non-key generating photons during QC. From the perspective of QKD operation, Eqs. (7a) and (7b) then directly yield the QBER of the system as

\begin{equation}
\begin{aligned}
QBER=\frac{1-\eta}{2},
\end{aligned}
\end{equation}
where the interference visibility is defined as $\eta =\left( P_{\text{R}}-P_{\text{T}} \right) /\left( P_{\text{R}}+P_{\text{T}} \right)= \cos \Delta \delta$.  
This result is formally analogous to that obtained in classical interference. Under quasi-static disturbances, such disturbances introduce only reciprocal phase shifts and therefore do not degrade the interference visibility of the system.

Our scheme employs a WM protocol featuring a large inherent time delay \cite{WM_our_work}. Assuming an initial optical power of $I_{\rm in}$, the output power after post-selection is denoted as $I_{\rm out}$. Taking the reflected port as an example, the output intensity satisfying $I_{\text{out}}^{\left( \text{R} \right)} = I_{\rm in} P_\text{R}$. 
 In the initial experimental stage, with no external disturbances applied to the Sagnac-loop, the time delay in the channel is solely due to birefringence, i.e., $\tau = \tau_0$. The post-selection angle is adjusted to minimize the output intensity at the reflected port, yielding a minimum value $I_{\text{min}}^{\left( \text{R} \right)}$ at an angle of $\epsilon_0$, 
\begin{equation}
\begin{aligned}
I_{\text{min}}^{\left( \text{R} \right)}=\frac{1}{4} I_{\rm in}\left(1+\cos \Delta \delta\right) \left( 1-e^{-\sigma^{2} \tau^{2}} \right),
\end{aligned}
\end{equation}
with $\omega_0 \tau_0 -\epsilon_0= k\pi, \ k = 0, 1, 2, \dots$ Starting from $I_{\rm min}$, the post-selection angle is finely adjusted by $\Delta \epsilon$, resulting in a post-selection angle of
$\epsilon = \epsilon_0 + \Delta \epsilon$, with the corresponding output intensity denoted as $I_{\text{1}}^{\left( \text{R} \right)}$:
\begin{equation}
\begin{aligned}
I_{\text{1}}^{\left( \text{R} \right)}=\frac{1}{4} I_{\rm in}\left(1+\cos \Delta \delta \right) \left[ 1-e^{-\sigma^{2} \tau^{2}}\cos \left( 2\Delta \epsilon \right) \right].
\end{aligned}
\end{equation}

Then an external disturbance is applied to the fiber, inducing a time delay variation $\tau=\tau_0+\Delta \tau$, and the resulting output intensity is denoted as $I_{\text{d}}^{\left( \text{R} \right)}$:
\begin{equation}
\begin{aligned}
I_{\text{d}}^{\left( \text{R} \right)}=\frac{1}{4} I_{\rm in}\left(1+\cos \Delta \delta \right) \left[ 1-e^{-\sigma^{2} \tau^{2}}\cos \left[ 2\left( \Delta \epsilon -\omega_{0} \Delta \tau \right) \right] \right],
\end{aligned}
\end{equation}
here, $\omega_0 \Delta \tau$ represents the RPD shift from the external disturbance. Under the approximations $\omega_{0} \Delta \tau \ll \Delta \epsilon \ll 1$ and $\sigma^{2} \tau^{2} \ll 1$, the intensity contrast ratio is given by:
\begin{equation}
\begin{aligned}
ICR =\frac{I_{\text{1}}^{\left( \text{R} \right)}-I_{\text{d}}^{\left( \text{R} \right)}}{I_{\text{1}}^{\left( \text{R} \right)}-I_{\text{min}}^{\left( \text{R} \right)}} \approx \left(1+\cos \Delta \delta \right)\frac{\omega_{0} \Delta \tau }{\Delta \epsilon}.
\end{aligned}
\end{equation}

In contrast, time-varying disturbances induce not only a RPD, but also an additional time-dependent GPD in the loop, such that the effective GPD becomes $\Delta \delta' = \Delta \delta + \Delta \delta_{\mathrm{d}}(t)$. These dynamic disturbances reduce the interference visibility and, consequently, directly degrade the communication performance, highlighting the need for timely identification and localization.

To address this, the NFL module exploits the nonreciprocal phase shift generated in a Sagnac interferometer. When an external disturbance of length $d$ (with $d \ll L$) occurs at an arbitrary position $x$ (where $x \neq L/2$) along a fiber loop of total length $L$, it perturbs the counter-propagating CW and CCW light waves at different times. This time delay results in a GPD between the two waves. The relationship between the disturbance position $x$ and the resulting phase shift can be derived by analyzing the propagation path difference. This relationship directly correlates the spectral characteristics of the interferometric output with the location of the disturbance, enabling precise localization \cite{ZF_ref}.

The total interference intensity $I$ results from the superposition of the two counter-propagating waves and is given by:
\begin{equation}
I = I_\text{CW} + I_\text{CCW} + I_\text{ac},
\end{equation}
where $I_\text{ac}$ is the cross-interference term:
\begin{equation}
I_\text{ac} = \left( e^{i\delta_\text{CW}}e^{-i\delta_\text{CCW}} + e^{-i\delta_\text{CW}}e^{i\delta_\text{CCW}} \right) \sqrt{I_\text{CW} I_\text{CCW}},
\end{equation}
here, $\delta_\text{CW}$ and $\delta_\text{CCW}$ represent the static global phases induced by the optical path length for the CW and CCW waves, respectively. Their difference is zero due to the common path of the Sagnac-loop.

As shown in the Sagnac-loop schematic, an input light wave of intensity $I_\text{in}$ is split using a CIR and a BS into CW and CCW components. Considering the reflected interference port, the output intensity can be expressed as
\begin{equation}
\begin{aligned}
 I_\text{R} = I_{\text{in}} \left( 1 + \cos \Delta \delta' \right),
\end{aligned}
\end{equation}
the total GPD between the two paths is defined as $\Delta \delta' =\Delta \delta_\text{d} \left( t \right) +\Delta \delta$, here, we neglect the static optical path difference between CW and CCW.
The term $\Delta \delta_\text{d}(t)$ corresponds to the time-varying phase shift induced by the external disturbance. Assuming that the disturbance has a spatial extent much smaller than the total fiber length $L$, the resulting phase drift can be expressed as:
\begin{equation}
\begin{aligned}
\Delta \delta_{\text{d}} \left( t \right) =\Delta \delta_{\text{d}_{\text{CW}}} \left( t \right) - \Delta \delta_{\text{d}_{\text{CCW}}} \left( t \right),
\end{aligned}
\end{equation}
where $ \Delta \delta_{\text{d}_{\text{CW}}} \left( t \right) =\Delta \delta_{\text{d}} \left( t \right)=\delta_\text{d} \sin \left( \omega_\text{s} t \right)$, $\delta_\text{d}$ and $\omega_\text{s}$ represent the amplitude and angular frequency of the phase drift, respectively. The resultant time-varying phase drift in the optical fiber is therefore given by:
\begin{equation}
\begin{aligned}
\Delta \delta' = \Delta \delta_{\text{d}} \left( t \right) - \Delta \delta_{\text{d}} \left( t-\frac{n}{c} \left( L-2x \right) \right) + \Delta \delta.
\end{aligned}
\end{equation}

Although both CW and CCW beams experience the disturbance at the same position $x$ in the loop, their counter-propagating nature causes them to interact with the disturbance at different times. Thus, $\Delta \delta_\text{d}(t)$ denotes the phase change induced in the CW beam at $x$, while $\Delta \delta_{\text{d}} \left( t - \frac{n}{c} (L - 2x) \right)$ represents the corresponding phase change in the CCW beam at the same location but at a distinct time instant, with $n$ and $c$ represent the refractive index and speed of light in the optical fiber.

With the path difference defined as $\Delta x = L - 2x$, the cross-interference component $I_\text{ac}$ of the interference intensity simplifies to:
\begin{equation}
\begin{aligned}
I_\text{ac}=-I_{0}\delta_\text{d} \cos \left[ \omega_\text{s} \left( t-\frac{nL}{2c} \right) \right] \sin \left( \frac{\omega_\text{s} n}{2c} \cdot \Delta x \right).
\end{aligned}
\end{equation}

As derived from the above equation, the amplitude of the $I_\text{ac}$ component is
$\text{AMP}_{I_{\text{ac}}} = -I_{0}\delta_\text{d} \sin \left( \frac{\omega_\text{s} n}{2c} \cdot \Delta x \right)$.
This amplitude vanishes when the sinusoidal argument satisfies \( \frac{\omega_\text{s} n}{2c} \cdot \Delta x = k\pi \), where \( k \) is a natural number. This condition defines a series of null frequencies in the interference spectrum. The \( k \)-th null frequency is given by $f_{\text{s, null}}=\frac{kc}{n\cdot \bigtriangleup x}$. This expression establishes the relationship between the disturbance location \( x \) and a measured null frequency \( f_\text{s, null} \) via the path difference $ \Delta x $, thereby enabling precise localization.
\begin{equation}
\begin{aligned}
x=\frac{1}{2} \left( L-\frac{kc}{nf_\text{s,null}} \right).
\end{aligned}
\end{equation}

The key to the localization system lies in ensuring that the CW and CCW light beams acquire distinct phase information from the disturbance. This requires that at least one of the optical pulses is passing through the disturbance point when it occurs, which imposes a condition on the pulse duration $\sigma_{t}$ relative to the disturbance length $d$ and the speed of light $c$, namely $\sigma_{t} \gg d/c$.

The SLIS uniquely enables simultaneous extraction of disturbance magnitude and location from events. Spectral analysis provides localization via the NFL module, while optical power measurements quantify disturbance strength. The independence of these analyses ensures parallel operation without cross-talk, yielding a robust multifunctional platform.

\section{Experimental setup}
\justifying
This section describes the experimental implementation of the SLIS, which integrates Sagnac-loop QKD with perception to enable highly stable key distribution and channel diagnosis without active compensation. The system accurately identifies both the magnitude and location of external disturbances along the fiber, thereby enhancing operational reliability. A schematic of the experimental setup is shown below.

\begin{figure}[htbp]
	\centering
	\includegraphics[width=0.6\textwidth]{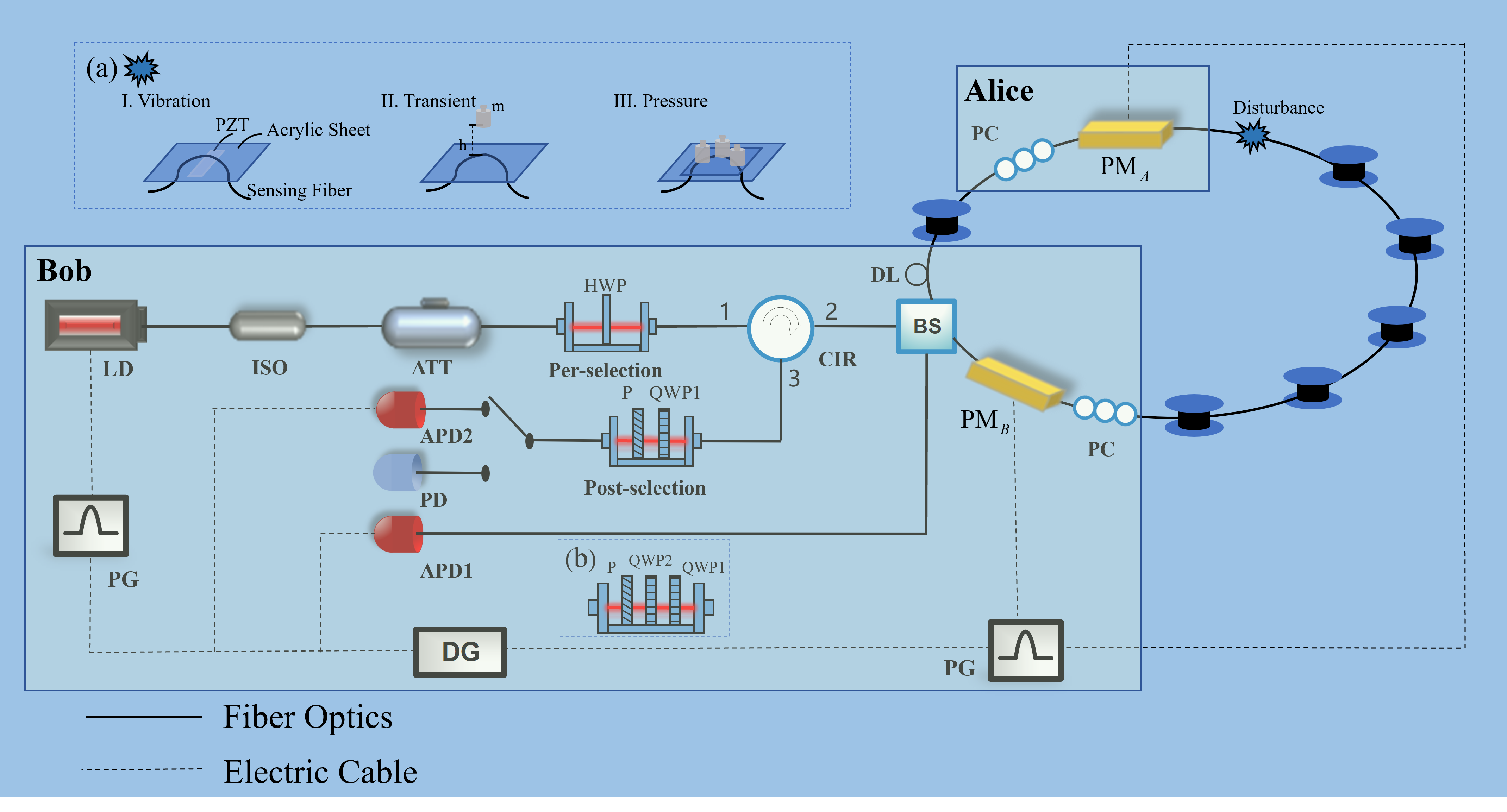}
	\caption{Schematic diagram of the SLIS. 
	(a) Application methods for three types of disturbances: I. PZT-induced vibrations, II. transient impacts, and III. quasi-static pressure.
(b) Post-selection setup for simulating channel time delay variations in WM sensing. }
	\label{SLIS}
\end{figure}

In the SLIS, a tunable single-frequency laser diode (LD, DenseLight) serves as the shared light source. The laser operates in continuous, digital, and analog modes, capable of generating single-frequency continuous or pulsed linearly polarized light with a central wavelength of 1550 $\text{nm}$ and a linewidth of 3.7 $\text{kHz}$. For the QKD module, the laser operates digitally to produce optical pulses with a repetition rate of 100 $\text{MHz}$ and a pulse duration of 2 $\text{ns}$. For sensing and localization, however, the system utilizes the continuous-wave operation of the same laser to meet the requirement for an extended interaction time between the optical beam and the disturbance in the localization module.

$\bf{Communication\  system}$, the pulsed light from the laser source first passes through an optical isolator (ISO, Thorlabs) to prevent back-reflected light originating in the Sagnac loop from re-entering the source, thereby protecting the laser. The pulse train is then attenuated by a variable optical attenuator (ATT, Linda) to precisely set its mean photon number to $\mu = 0.1$. This ensures that the output pulses are in the weak coherent state regime, effectively emulating a single-photon source and providing a secure, stable foundation for QKD.

Subsequently, the pulsed light then enters the Sagnac-loop through a fiber circulator (CIR, Thorlabs) and is split by a beam splitter (BS, ORTE Photonics) into two counter-propagating components: the CW and CCW pulses. The total length of the Sagnac-loop is 30 km. Inside the Sagnac-loop, the pulse exiting the BS passes through a polarization controller (PC, Thorlabs) for polarization adjustment to optimize the interference visibility. Following this adjustment, the pulse is coupled into a fiber delay line (DL, OZ Optics), which fine-tunes the total fiber length to ensure a temporal offset between the CW and CCW pulses at a specific position. This temporal separation enables a time-division phase modulation scheme. 

A phase modulator (PM, Conquer) with a half-wave voltage of $V_{\pi} = 3.2~\text{V}$, a modulation pulse width of $\Delta t_\text{PM}=3.3\ \text{ns}$, and a modulation frequency $f_\text{PM}=100\ \text{MHz}$ is placed near the BS, serving as Bob’s modulator. By precisely controlling the timing of its electrical drive signal, this PM independently modulates the CW pulse to encode Bob's information. A second, identical PM is positioned at a non-central point in the loop, serving as Alice’s modulator. It independently modulates the CCW pulse in the same manner to encode Alice's information.

After traversing the entire optical path in the loop, the CW and CCW pulses are independently modulated by the respective PMs of the two communicating parties. Upon returning to the BS, the two pulses interfere with each other. The interference outputs at the transmission and reflection ports are detected by single-photon avalanche photodiodes (APD, Qasky), which operate with a gate width of $\Delta t_{\text{APD}}=2\  \text{ns}$, a repetition frequency of $f_{\text{APD}}=100\ \text{MHz}$, and a maximum detection efficiency of 20 $\%$.

\begin{figure}[ht]
    \centering
    \hspace{-0.4cm} 
    \begin{tabular}{c@{\hspace{-0.1cm}}c} 
        \begin{tikzpicture}
            \node[anchor=south west,inner sep=0] (image1) at (0,0) {\includegraphics[width=5.4cm,height=4.6cm]{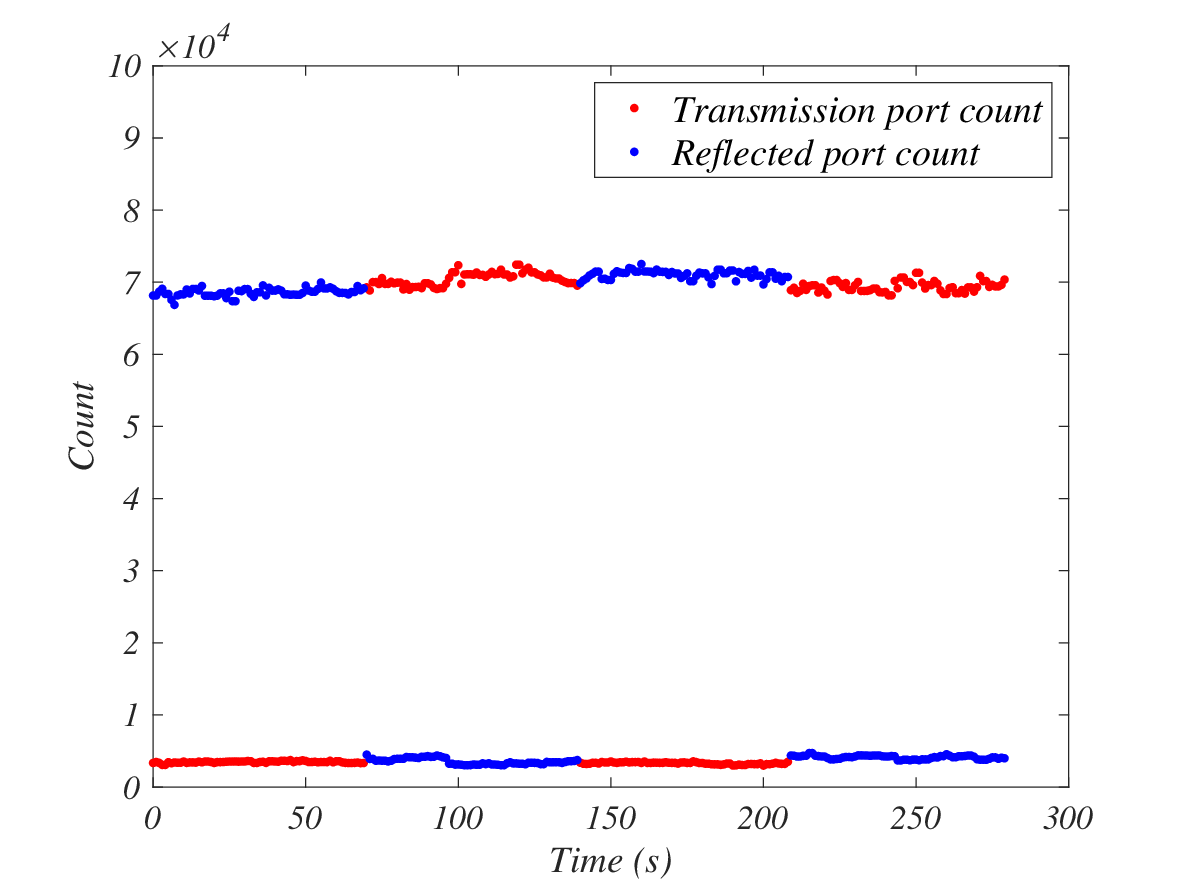}};
            \node at (0.1,4.4) {\textbf{(a)}};
        \end{tikzpicture} &
        \begin{tikzpicture}
            \node[anchor=south west,inner sep=0] (image2) at (0,0) {\includegraphics[width=5.4cm,height=4.6cm]{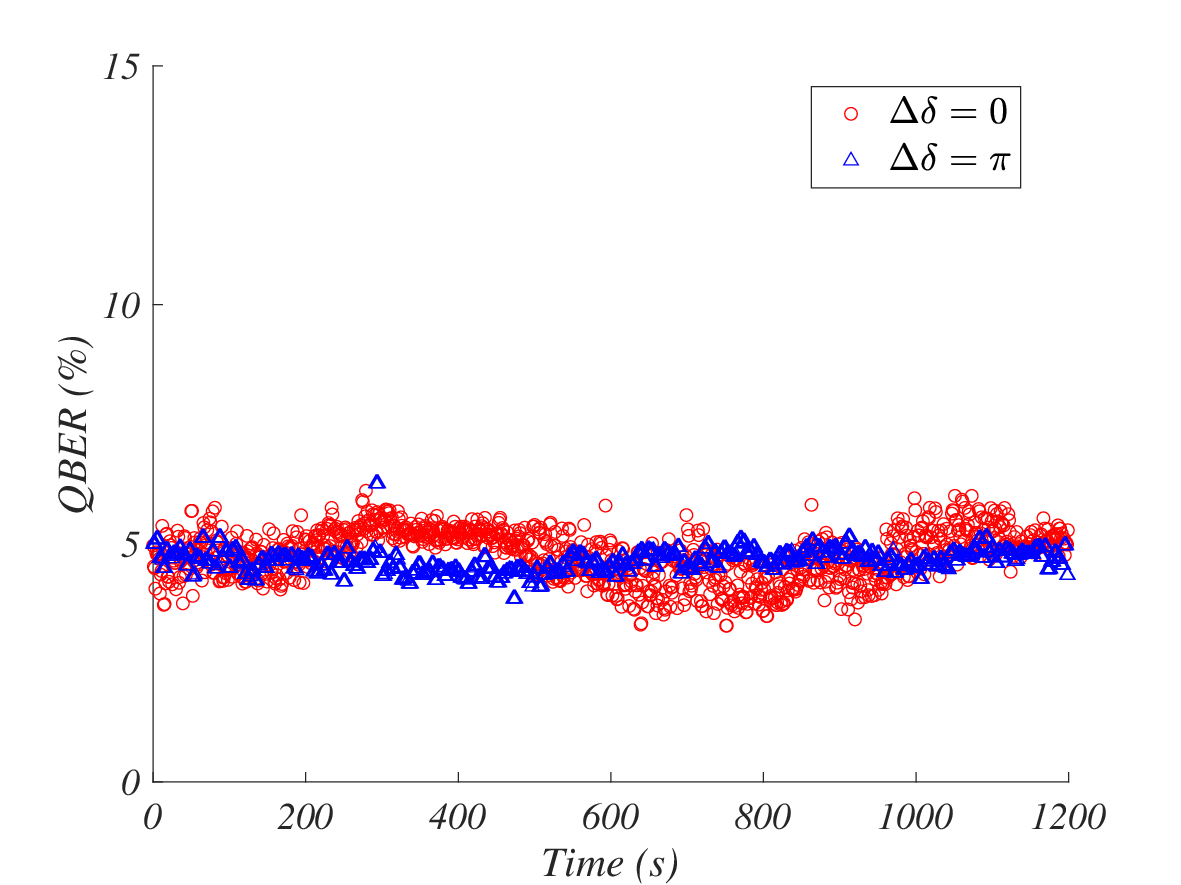}};
            \node at (0.1,4.4) {\textbf{(b)}};
        \end{tikzpicture} \\
    \end{tabular}
    \caption{Experimental results of QKD performance in the SLIS.}
    \label{QKD}
\end{figure}

In the experiment, $\text{PM}_A$ and $\text{PM}_B$ were placed at asymmetric positions along the 30 $\text{km}$ Sagnac-loop. When different voltages ($V_{0}$, $V_{\pi/2}$, $V_{\pi}$, and $V_{3\pi/2}$) were applied to $\text{PM}_A$ and $\text{PM}_B$, corresponding phase shifts of $0$, $\pi/2$, $\pi$, and $3\pi/2$ were introduced, respectively. The photon counts at the transmission and reflection ports varied with the applied voltages, as illustrated in the Figure~\ref{QKD} (a).

The experimental results show that, even without any active phase stabilization, the Sagnac-loop based QKD module maintained an average QBER of 4.76$\%$ for a 0-phase difference and 4.63$\%$ for a $\pi$-phase difference over 20 minutes of continuous operation. This performance demonstrates the system's long-term stability in the absence of significant external disturbances. Under these conditions, the raw key generation rate was 22.4 $\text{kbps}$. In this module, the pulse output beam from the LD over 15 minutes, the light source power fluctuation is 0.19$\%$.

$\bf{Perception\  system.}$ When the QBER exceeds a predefined threshold or fiber-link monitoring is required, the system switches to the fiber sensing mode. In this mode, the LD is reconfigured from the digital modulation used for QKD to continuous-wave modulation, providing the extended optical duration required by the localization module. Simultaneously, the detection unit is switched from an APD to a photodetector (PD, Teledyne LeCroy Inc.). In this regime, the laser emits a continuous-wave beam with a central wavelength of 1550 nm, a linewidth of 3.7 kHz, and an output power of 5.645 mW.

The optical beam then passes through the ATT with its attenuation set to 0 dB, passes through the optical ISO, and enters the pre-selection polarization control module (FiberBench, FB, Thorlabs). This module contains a half-wave plate (HWP) to rotate the linearly polarized input light from source to an arbitrary orientation, thereby preparing the required initial polarization state.

In the experimental investigation, three representative types of disturbances in the Sagnac-loop are systematically examined, collectively covering the majority of practical scenarios encountered in real applications.

The first type is periodic vibrations induced by a PZT with a width of 2 cm. As illustrated in Figure~\ref{SLIS}(a)I, the PZT element is clamped together with a rigid support onto a bare fiber segment positioned between two fiber spools, thereby applying a mechanical vibration at a fixed frequency described by $V = V_0 \sin(\omega_s t) $, where $V_0$ denotes the driving voltage amplitude and $\omega_s$ is the frequency. Such rapidly time-varying disturbances cause the CW and CCW propagating waves to experience different vibration phases, thereby breaking the intrinsic self-compensation property of the Sagnac-loop. The phase shift induced by the PZT can be expressed as:
\begin{equation}
\Delta \delta_{\text{d}_{\text{PZT}}} \left( t \right) = \delta_{\text{d}_{\text{PZT}}} \sin \left( \omega_{s} t \right),\  \delta_{\text{d}_{\text{PZT}}} \propto \frac{2\pi}{\lambda} CEd_{33}V_{0}L,
\end{equation}
where $C$ is the stress–optic coefficient, $E$ is the Young’s modulus of the fiber, and $d_{33}$ denotes the piezoelectric coefficient of the PZT. This expression establishes a direct phase-transfer model linking the applied PZT driving voltage to the induced optical phase shift in the bare fiber segment. The same theoretical framework naturally extends to disturbance localization.

The second type is transient impact, realized by dropping a mass $m$ from a fixed height $h$ onto a bare fiber section within the loop. Compared to the overall fiber length, the impacted region can be approximated as a point-like source. The resulting phase shift $\Delta \delta_{\text{d}_\text{TRA}}\left( t \right)$ is governed by the photoelastic effect and primarily arises from a transient refractive-index variation $\Delta n$. Owing to the localized and ultrashort nature of the impact, its spatiotemporal distribution can be approximated by a Dirac delta function $\mathbb{D}(x - x_0, t - t_0)$, where $x_0$ and $t_0$ denote the impact position and time, respectively, as shown in Figure~\ref{SLIS}(a)II.  The corresponding phase shift is given by:
\begin{equation}
\begin{aligned}
\Delta \delta_{\text{d}_\text{TRA}}\left( t \right) =\Delta \delta_{\text{d}_\text{TRA}}\mathbb{D} \left( t-t_{0}, x-x_{0} \right),\  \Delta \delta_{\text{d}_\text{TRA}} \propto \frac{2\pi}{\lambda} \cdot \left( n_\text{eff}+C \right) \cdot \chi \cdot \frac{\sqrt{2gh} \cdot m}{S\cdot E} .
\end{aligned}
\end{equation}
where $\lambda$ is the operating wavelength, $ n_{\mathrm{eff}} $ denotes the effective refractive index, $ C$ is the stress–optic coefficient, $E$ is the Young’s modulus of the fiber, $\chi$ represents the coupling efficiency between the impact force and the fiber core, $S$ is the contact area, and $g$ is the gravitational acceleration. The Dirac delta function $\delta(t - t_0, x - x_0)$localizes the perturbation in both space and time, thereby characterizing the spatiotemporal distribution of the impulsive force along the fiber.

\begin{figure}[ht]
    \centering
    \hspace{-0.4cm} 
    \begin{tabular}{c@{\hspace{-0.1cm}}c} 
        \begin{tikzpicture}
            \node[anchor=south west,inner sep=0] (image1) at (0,0) {\includegraphics[width=5.4cm,height=3.4cm]{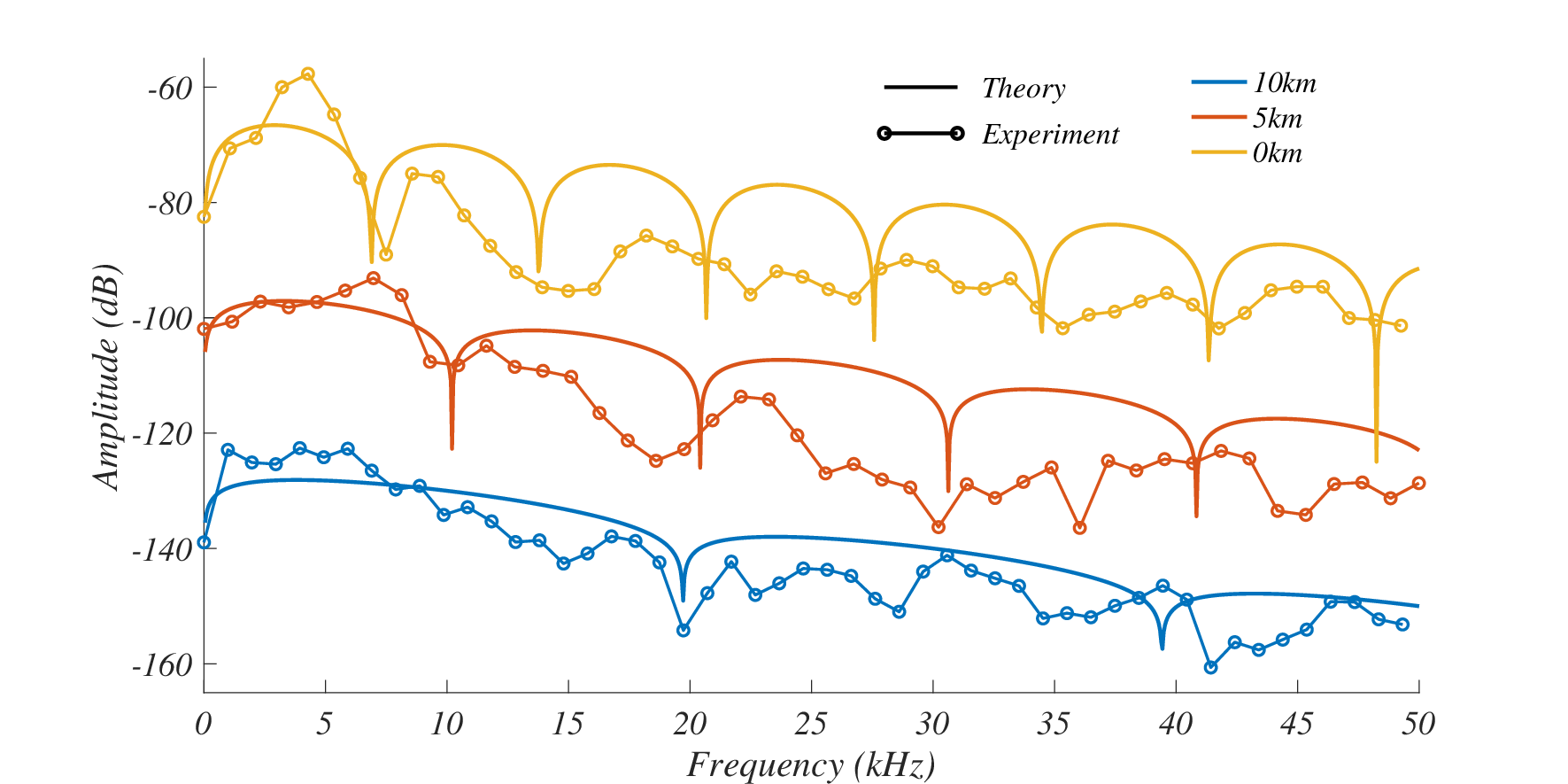}};
            \node at (0.2,3.2) {\textbf{(a)}};
        \end{tikzpicture} &
        \begin{tikzpicture}
            \node[anchor=south west,inner sep=0] (image2) at (0,0) {\includegraphics[width=5.6cm,height=3.4cm]{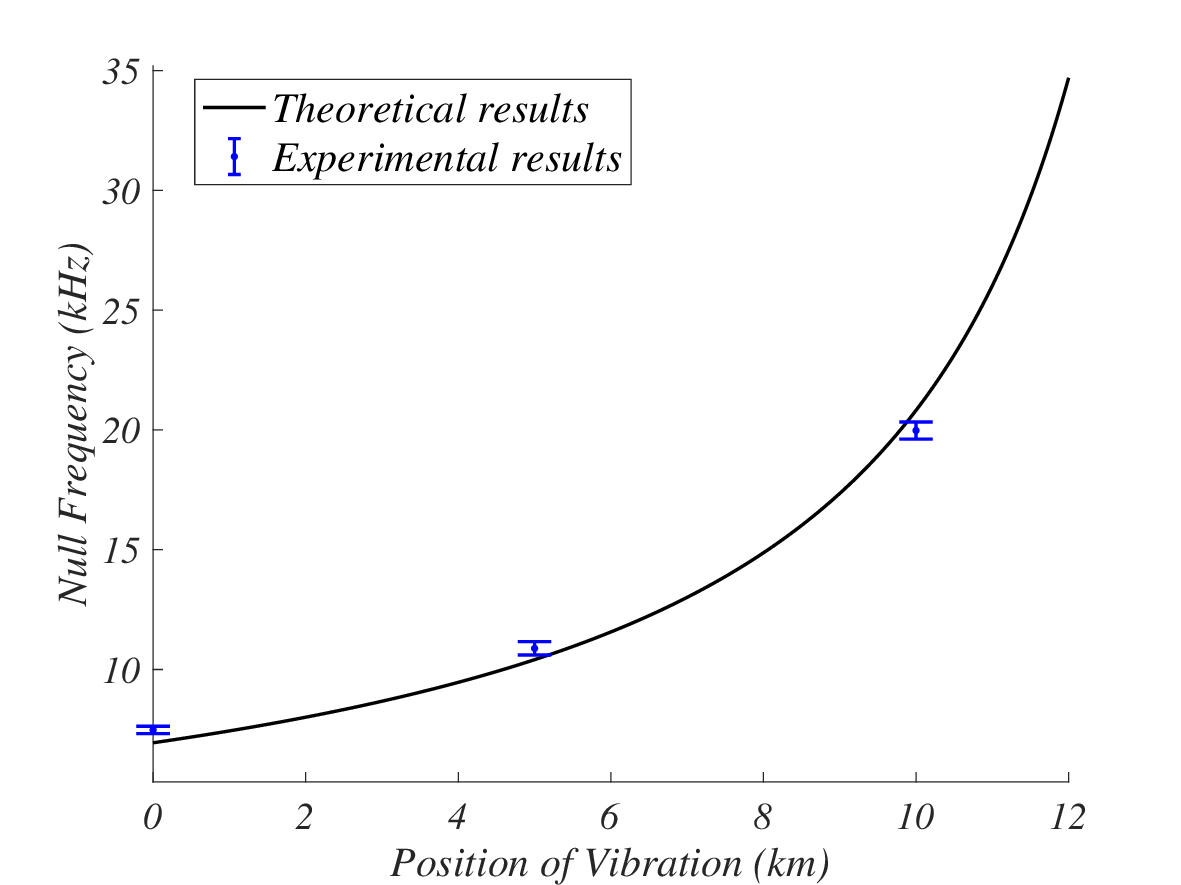}};
            \node at (0.2,3.2) {\textbf{(b)}};
        \end{tikzpicture} \\
                \begin{tikzpicture}
            \node[anchor=south west,inner sep=0] (image3) at (0,0) {\includegraphics[width=5.4cm,height=3.4cm]{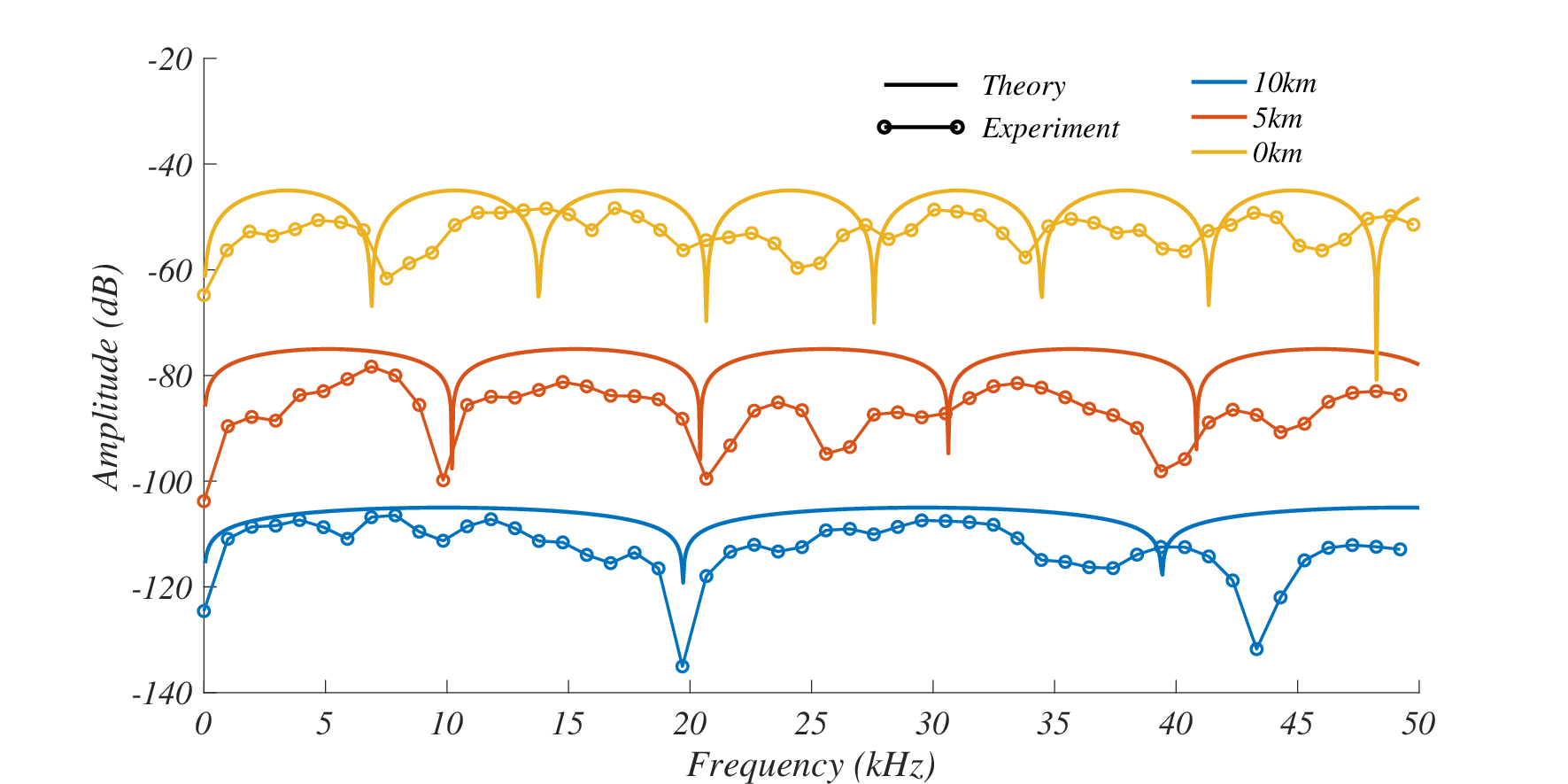}};
            \node at (0.2,3.2) {\textbf{(c)}};
        \end{tikzpicture} &
        \begin{tikzpicture}
            \node[anchor=south west,inner sep=0] (image4) at (0,0) {\includegraphics[width=5.6cm,height=3.4cm]{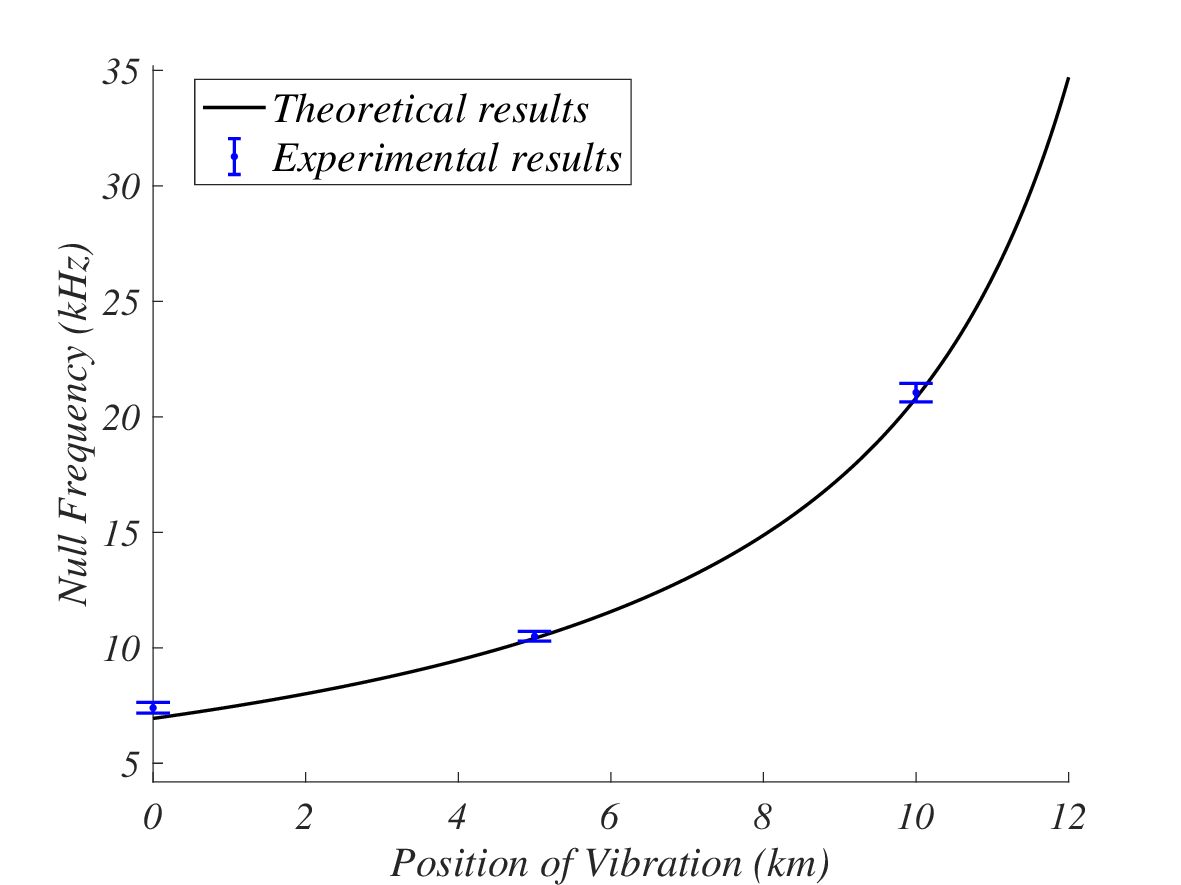}};
            \node at (0.2,3.2) {\textbf{(d)}};
        \end{tikzpicture}
    \end{tabular}
    \caption{Localization spectra and error analysis for two disturbance types at different positions in the 30 $\text{km}$ loop. (a) Localization spectra for periodic vibrations applied by the PZT at different positions. (b) Statistical error analysis from experiments where the PZT vibration was applied at the same position but with different frequencies and amplitudes. (c) Localization spectra for transient impacts applied at different positions. (d) Statistical error analysis from experiments where the impact was applied at the same position using different masses.}
    \label{Localization}
\end{figure}
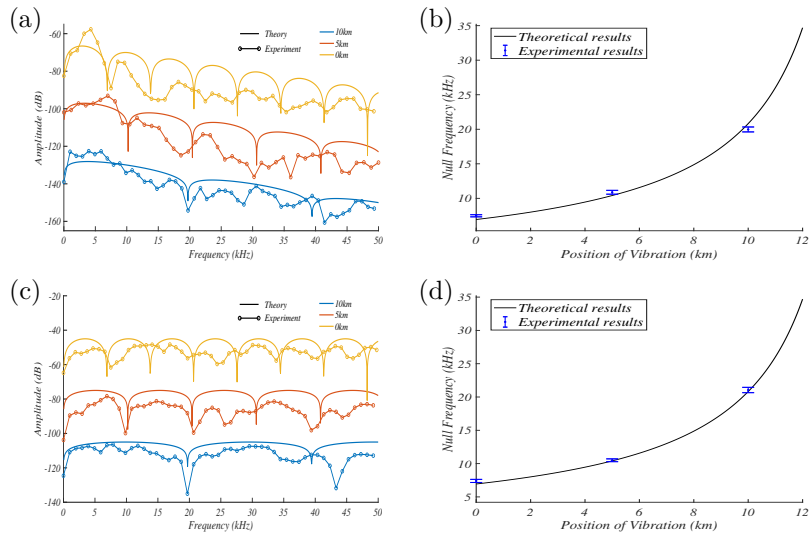

To suppress mechanical and acoustic noise induced by the disturbance sources, the external disturbance units were enclosed in a sponge–soundproofing material–sponge sandwich box with a volume of approximately 0.045 $\text{m}^3$, through which a 20 $\text{cm}$ bare fiber segment was routed. Experiments on the two types of dynamic disturbances were performed under these noise-controlled conditions, with the results shown in Figure~\ref{Localization}. Additional results on QKD stability and localization performance under different parameter settings are provided in the Supporting Information S2.

Experimental results demonstrate the localization performance of the SLIS under external disturbances is shown in Table~\ref{tab:NFL_perfor}. The localization resolution is fundamentally limited by the frequency resolution of the photodetector's oscilloscope, as described by 
\begin{equation}
\begin{aligned}
R_{s}=\frac{kc}{n} \left| \frac{\Delta f}{f_{\text{null, s}}^{2}-\left( \Delta f \right)^{2}} \right|, 
\end{aligned}
\end{equation}
with \( \Delta f = 0.5 \, \text{kHz} \) being resolution limit of the instrument. This relationship results in a position-dependent resolution, spanning from \( 0.02 \, \text{m} \) at the nearest point to \( 542.2 \, \text{m} \) at the farthest point in the loop. The overall localization accuracy, evaluated over all recorded disturbance events, exhibits a mean error of \( 114.2 \, \text{m} \). 

To ensure a comprehensive evaluation, the localization error is quantified by comparing the estimated and true disturbance positions across all recorded events. The error is evaluated through error propagation:
\begin{equation}
\sigma_x = \left|\frac{dx}{df}\right| \sigma_f,
\end{equation}
where \( \sigma_f \) denotes the uncertainty in frequency measurement. By applying this formula with the standard deviation of the measured frequencies from all experimental samples, we obtain the position-dependent localization error, which varies with propagation distance, as summarized in Table~\ref{tab:NFL_perfor}.

\begin{table}[!t]
\centering
\caption{Average standard deviation of intensity contrast ratio for sensing fibres of length 1 m and 10 m at different post-selection angles}
\label{tab:NFL_perfor}
\begin{tabular}{|c|c|c|c|}
\hline
\textbf{Parameter}                       & \textbf{0 km}                            & \textbf{5 km}                        & \textbf{10 km} \\
\hline
Theoretical NF Point                   & $6.91\  \text{kHz}$                     & $10.21\  \text{kHz}$           & $19.72\  \text{kHz}$ \\  
\hline
Theoretical Resolution                & $0.02\  \text{m}$                       & $125.02\  \text{m}$              & $500.31\  \text{m}$ \\  
\hline
Error @ vibration                         & $286.83\  \text{m}$                   & $244.84\  \text{m}$            & $93.04\  \text{m}$ \\  
\hline
Error @ transient                         & $441.27\  \text{m}$                   & $199.12\  \text{m}$           & $94.18\  \text{m}$ \\  
\hline
\end{tabular}
\end{table}

The response of the SLIS to a disturbance critically depends on its temporal dynamics, governed by the interferometer's reciprocal nature. For dynamic disturbances, such as the periodic vibrations and transient impacts discussed above, the induced phase shift is non-reciprocal and time-varying. The counter-propagating pulses experience different phase shifts when they traverse the disturbance point at different times, leading to a significant non-reciprocal phase difference. This GPD shift can be effectively detected using conventional interferometry.

In contrast, for the third type of quasi-static disturbance—such as sustained pressure applied by standard weights that slowly modifies the fiber birefringence—the CW and CCW pulses acquire identical phase shifts, resulting in a vanishing net GPD. Nevertheless, information about the quasi-static disturbance is encoded in the RPD between orthogonal polarization components. To access this information with high sensitivity, we employ a fiber-based WM scheme. Supporting Information S3 demonstrates the high sensitivity of WM enhanced sensing. Specifically, the WVA effect under different post-selection angles is investigated by simulating RPD variations using wave plates. The configuration of the post-selection module is illustrated in Figure~\ref{SLIS}(b).

To further validate the capability of the integrated system for sensing quasi-static disturbances, a pressure-sensing experiment was performed. As illustrated in Figure~\ref{SLIS}(a)III, a standard weight was placed on a $20~\mathrm{cm}$ section of bare fiber. The post-selection angle was fixed at $\Delta\epsilon = 30^\circ$, and the applied load was increased stepwise in increments of $100~\mathrm{g}$. Each weight was carefully centered and placed upright to ensure uniform pressure distribution along the fiber segment.

\begin{figure}[ht]
    \centering
    \hspace{-0.4cm} 
    \begin{tabular}{c@{\hspace{-0.1cm}}c} 
        \begin{tikzpicture}
            \node[anchor=south west,inner sep=0] (image1) at (0,0) {\includegraphics[width=5cm,height=4cm]{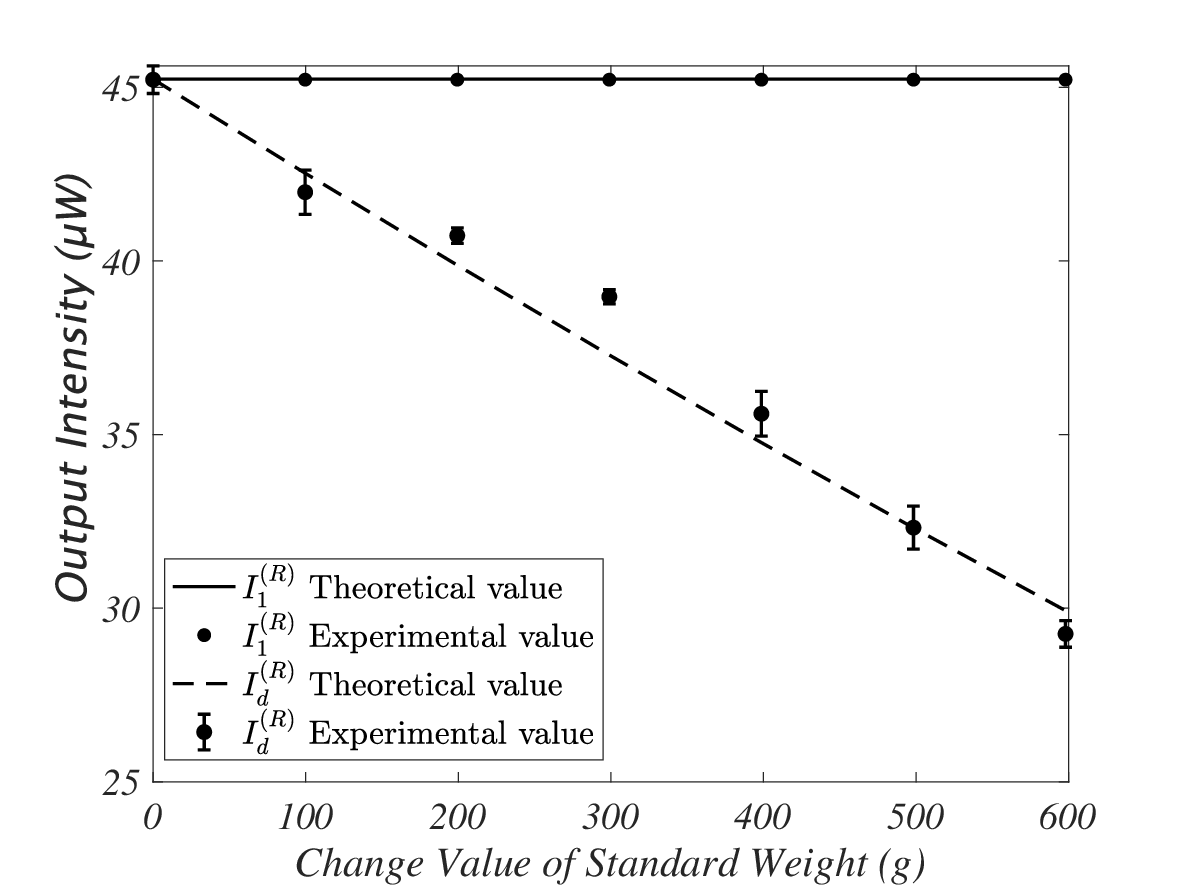}};
            \node at (0.2,3.6) {\textbf{(a)}};
        \end{tikzpicture} &
        \begin{tikzpicture}
            \node[anchor=south west,inner sep=0] (image2) at (0,0) {\includegraphics[width=5cm,height=4cm]{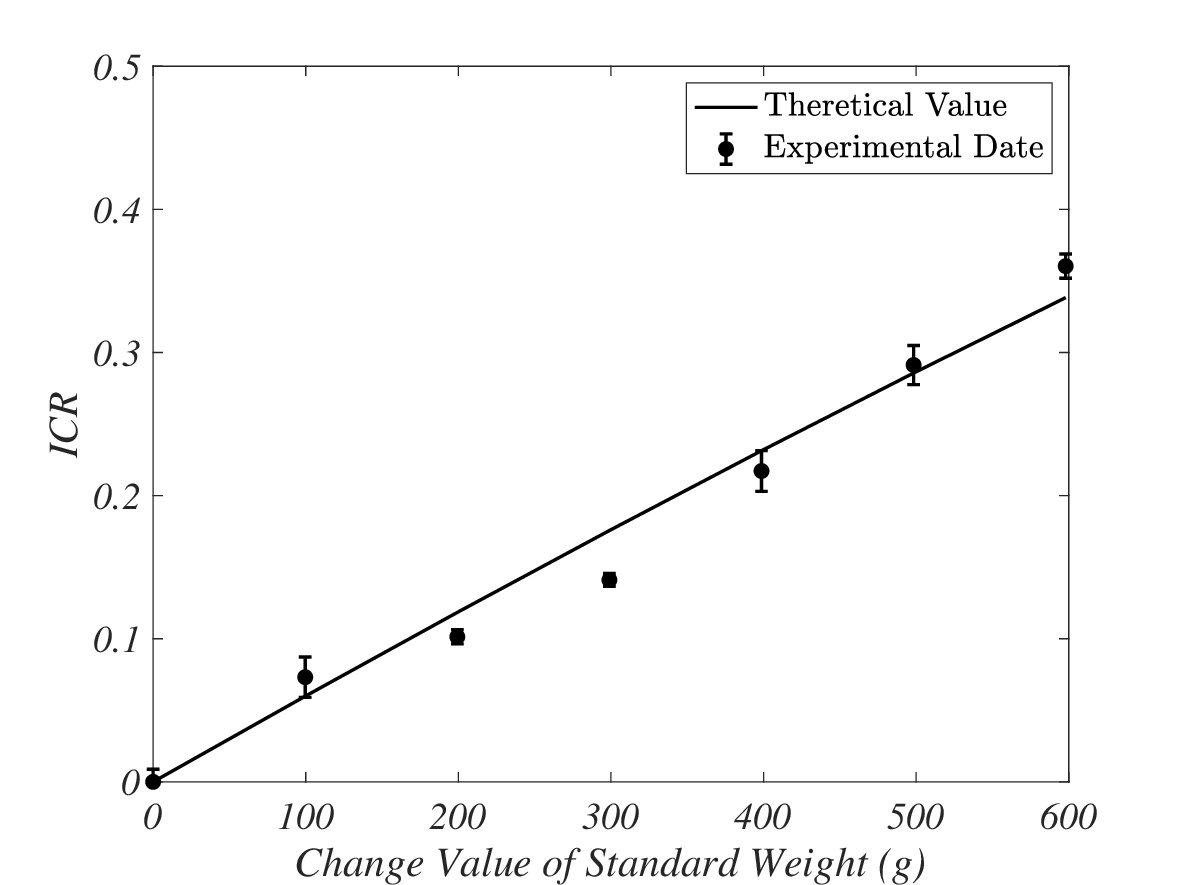}};
            \node at (0.2,3.6) {\textbf{(b)}};
        \end{tikzpicture} \\
                \begin{tikzpicture}
            \node[anchor=south west,inner sep=0] (image3) at (0,0) {\includegraphics[width=5cm,height=4cm]{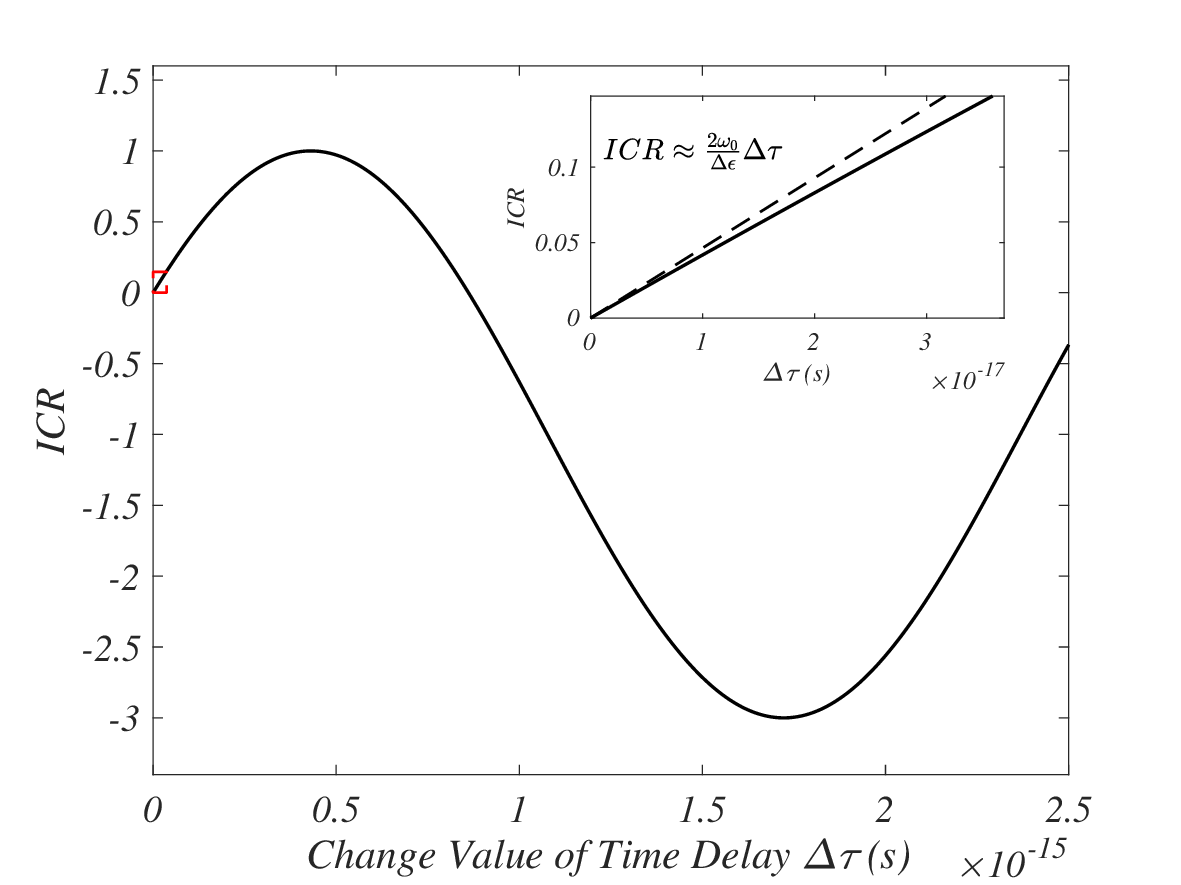}};
            \node at (0.2,3.6) {\textbf{(c)}};
        \end{tikzpicture} &
        \begin{tikzpicture}
            \node[anchor=south west,inner sep=0] (image4) at (0,0) {\includegraphics[width=5cm,height=4cm]{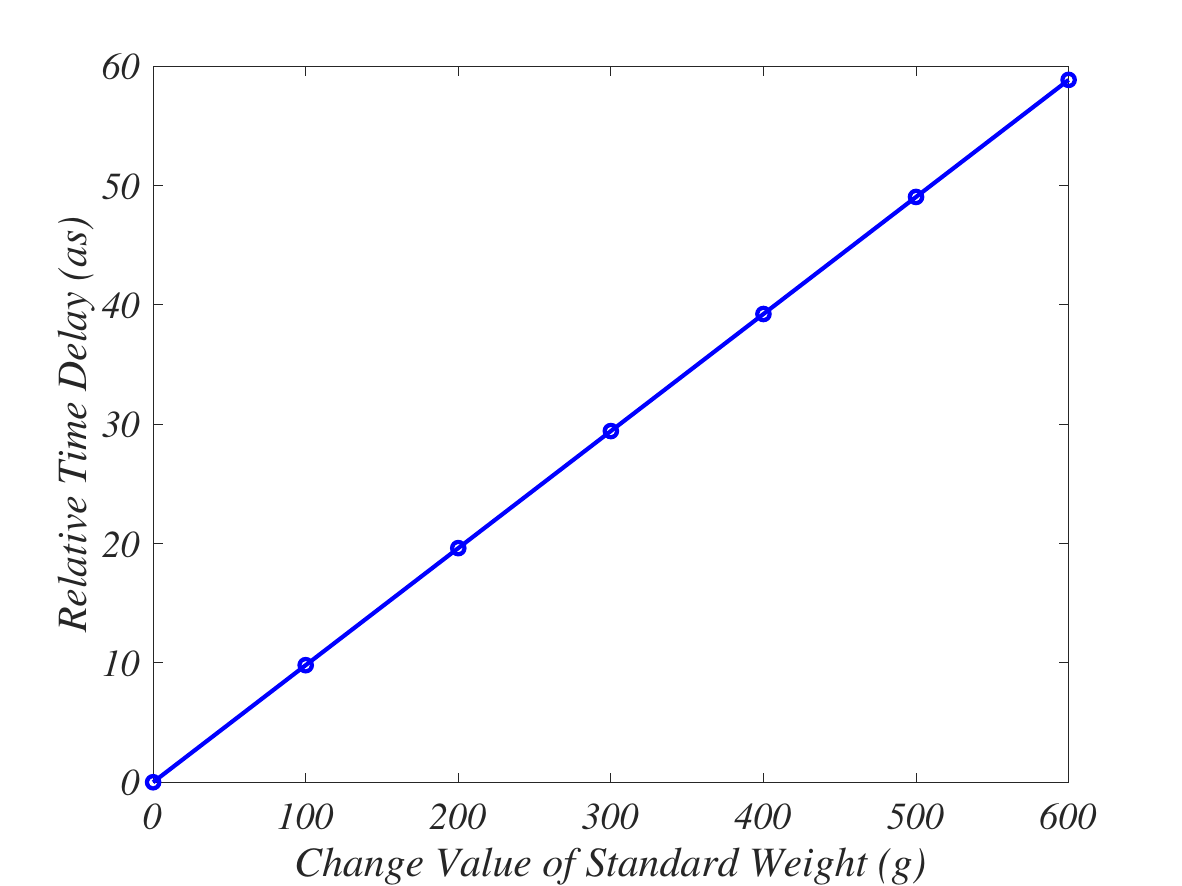}};
            \node at (0.2,3.6) {\textbf{(d)}};
        \end{tikzpicture}
    \end{tabular}
    \caption{Experimental results of quasi-static pressure sensing in the 30 km loop. (a) Theoretical and experimental values of the post-selected intensity $I_{\text{d}}^{\left( \text{R} \right)}$ versus applied mass. (b) ICR as a function of weights, derived from measured intensities. (c) Theoretical relationship between ICR and relative time delay at a post-selection angle $\Delta \epsilon$ of 30 $^\circ$. (d) Theoretical curve of relative time delay induced per 100 g mass increment on the bare fiber.}
    \label{WM}
\end{figure}
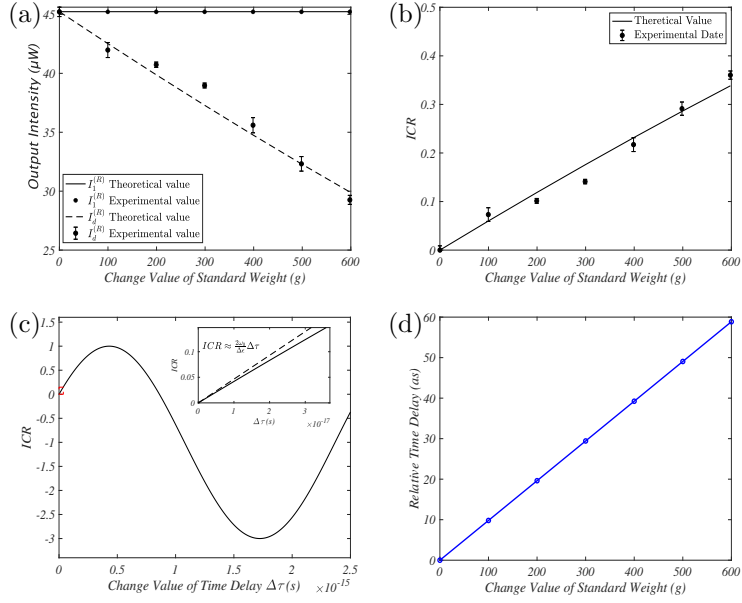

Figure~\ref{WM}(a) shows the corresponding post-selected intensity $I_{\text{d}}^{\left( \text{R} \right)}$ for each loading step. The extracted ICR as a function of the applied mass is presented in Figure~\ref{WM}(b), together with the theoretical prediction. Overall, the experimental results agree well with theory, while minor deviations at certain load levels are discussed later.

To clarify the correspondence between the applied mass and the induced relative time delay, Figure~\ref{WM}(c) presents the theoretical dependence of the ICR on the relative delay $\Delta\tau$ at $\Delta\epsilon = 30^\circ$. The inset highlights the small-delay regime ($0$–$10^{-17}~\mathrm{s}$), where the dashed curve represents the small-$\Delta\tau$ approximation given by Eq.~(11).

Figure~\ref{WM}(d) displays the theoretical relationship between the relative time delay and the applied weights, derived from the stress-induced birefringence effect in the SMF. The simplified theoretical expression is given by $\Delta \tau = C m g / (S c)$, where $C$ is the stress-optic coefficient with a theoretical value of $3 \times 10^{-12} \text{m/N}$, and $S$ is the effective area over which the force acts on the bare fiber, approximately $10^{-4} \text{m}^2$. According to this model, a 100 $\text{g}$ mass acting on a 10 $\text{cm}$ bare fiber section theoretically induces a relative time delay of 9.81 $\text{as,}$ which is consistent with the experimental results obtained from the WM sensing.

The primary factor affecting the results is polarization drift induced by the disturbance, as analyzed in Supporting Information S3. The single-mode sensing fiber, integrated with the channel fiber, lacks polarization-maintaining capability. Sustained pressure alters both the orientation and magnitude of the fiber's effective birefringence axis, modifying the beam's polarization and phase. To stabilize the polarization state, we implemented real-time polarization compensation via optical power feedback in the pre-selection module after each weight change.

It is worth emphasizing that the proposed WM enhanced sensing strategy is not limited to detecting slowly varying pressure-type disturbances. Since the sensing mechanism exploits the polarization degree of freedom of the optical field, it primarily measures the time-delay variations between two orthogonal polarization components, which are closely related to the birefringent properties of the fiber. Consequently, depending on the material characteristics of the employed optical fiber, this method can be extended—through appropriate calibration—to measure other parameters and adapt to diverse application scenarios. These include physical quantities such as hydrostatic pressure and temperature, or any external factors that induce variations in fiber birefringence, As depicted by the blue dashed line in Figure~\ref{WP}.     

To further demonstrate the comprehensive performance of the proposed SLIS, a comparative analysis is conducted with recently reported QKD–sensing integrated schemes. The comparison primarily focuses on the functional performance of each system, and the key evaluation metrics are summarized in Table~\ref{tab:qkd_table}.

\captionsetup[table]{justification=raggedright, singlelinecheck=false} 
\begin{table}[htbp]
  \centering
  \caption{Comparison of reported QKD-integrated sensing systems, including key rate and transmission distance for communication, as well as precision (\textbf{Pre}), resolution (\textbf{Res}), sensitivity (\textbf{Sen}), localization error (\textbf{Err}), and accuracy (\textbf{Acc}) for sensing.}
  \label{tab:qkd_table}
  \includegraphics[width=\linewidth]{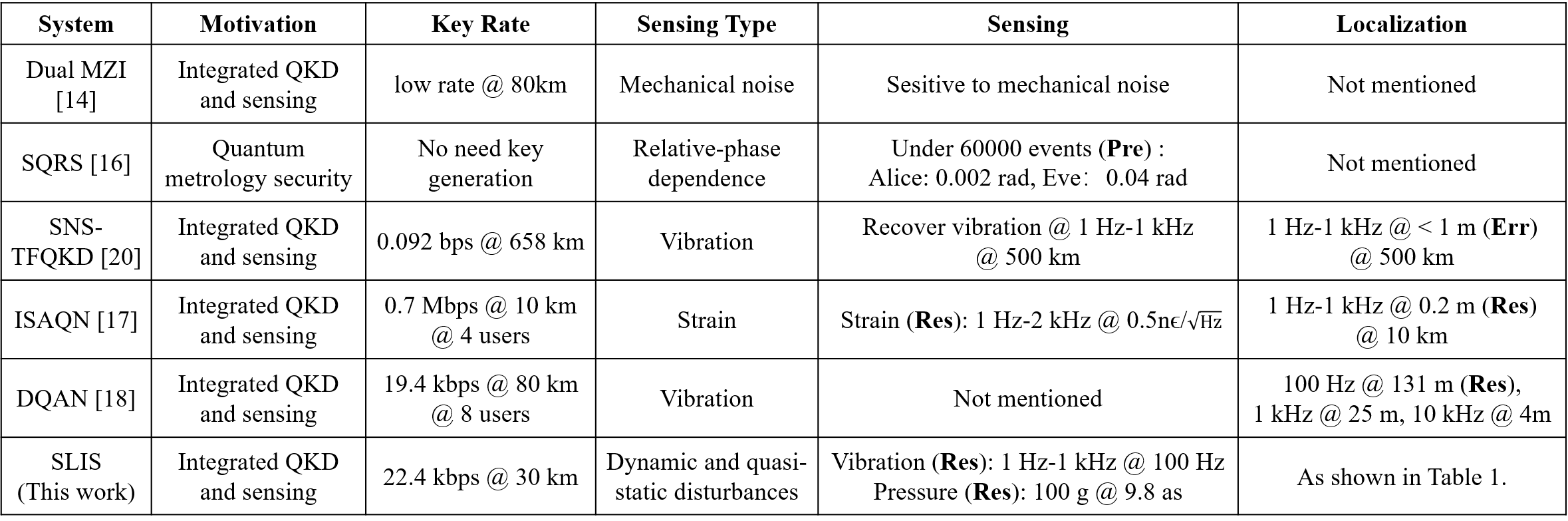}
\end{table}

Table~\ref{tab:qkd_table} analysis indicates that the proposed SLIS offers more comprehensive functionality than previously reported integrated systems, combining secure QC with both sensing and localization of external disturbances. The system supports a wider range of disturbance detection: dynamic, large-scale disturbances are monitored via interferometric measurement, while quasi-static, small-scale variations are captured through WM, significantly enhancing the scope and resolution of real-time channel disturbance monitoring.

From a system architecture perspective, communication and sensing operate in a time-division manner \cite{WM_QKD_signal}. Sensing is activated only when communication security is compromised, ensuring that the introduction of sensing does not introduce additional security risks. The sensing module thus serves as an auxiliary capability, endowing the communication system with environmental awareness and adaptive response, thereby improving overall robustness.

Regarding performance metrics, the QKD module, based on a bidirectional loop, exhibits limited transmission distance and leaves room for key rate optimization; however, its loop structure facilitates multi-user integration into a Sagnac-loop network without affecting sensing functionality. For perception, dynamic disturbances are detected with moderate precision due to added noise and detector resolution, whereas quasi-static, weak disturbances are captured with high sensitivity, demonstrating the system’s unique advantage. Overall, SLIS achieves high integration of communication and perception, providing enhanced adaptability to environmental disturbances.

\section{Conclusion}
In conclusion, we have theoretically and experimentally demonstrated a SLIS that enables environmental disturbance sensing and localization within a QKD framework. Operating in a time-division manner, the system switches from secure key distribution to perception when communication integrity is compromised, thereby enhancing robustness against environmental disturbances. Over a 30 km Sagnac loop, the SLIS achieves a stable key generation rate of 22.4 kbps with a QBER below 5$\%$ without active phase compensation. The system exhibits strong perception performance for both dynamic and quasi-static disturbances, enabling vibration detection down to 100 Hz with long-distance localization capability and attosecond-level time-delay sensitivity via WM enhancement. Moreover, by exploiting distinct degrees of freedom for communication and sensing, the SLIS supports wavelength-multiplexed concurrent operation, allowing real-time monitoring of channel conditions and potential eavesdropping activities, and offers multi-user scalability. These results establish SLIS as a practical and extensible platform for integrated QC and sensing, pointing toward self-diagnosing and resilient QN.

\medskip
\section*{\textbf{Acknowledgements}} \par 
This work is supported by National Natural Science Foundation of China (62371199, 62071186); Natural Science Foundation of Guangdong Province (2024A1515012427), Guangdong Provincial Quantum Science Strategic Initiative (GDZX2305001), Guangdong Provincial Key Laboratory (2020B1212060066). 

\medskip

\bibliographystyle{MSP}
\bibliography{Refference}

@article{WM_QKD_signal,
	author = {Song, Qi and Li, Hongjing and Huang, Jingzheng and Huang, Peng and Tan, Xiaorui and Tao, Yu and Shi, Chunhui and Zeng, Guihua},
	journal = {Communications Physics},
	number = {1},
	pages = {370},
	title = {Weak signal extraction in non-stationary channel with weak measurement},
	volume = {6},
	year = {2023}}

@article{SenCommu_integrated_overview,
	author = {Udd, Eric},
	doi = {10.1063/1.1145411},
	eprint = {https://pubs.aip.org/aip/rsi/article-pdf/66/8/4015/19216118/4015_1_online.pdf},
	issn = {0034-6748},
	journal = {Review of Scientific Instruments},
	month = {08},
	number = {8},
	pages = {4015-4030},
	title = {An overview of fiber‐optic sensors},
	url = {https://doi.org/10.1063/1.1145411},
	volume = {66},
	year = {1995}}

@article{SenCommu_integrated_highspeed,
	author = {Huang, Ming-Fang and Salemi, Milad and Chen, Yuheng and Zhao, Jingnan and Xia, Tiejun J. and Wellbrock, Glenn A. and Huang, Yue-Kai and Milione, Giovanni and Ip, Ezra and Ji, Philip and Wang, Ting and Aono, Yoshiaki},
	doi = {10.1109/JLT.2019.2935422},
	journal = {Journal of Lightwave Technology},
	number = {1},
	pages = {75-81},
	title = {First Field Trial of Distributed Fiber Optical Sensing and High-Speed Communication Over an Operational Telecom Network},
	volume = {38},
	year = {2020}}

@article{SenCommu_integrated_Light,
	author = {He, Haijun and Jiang, Lin and Pan, Yan and Yi, Anlin and Zou, Xihua and Pan, wb and Willner, Alan and Fan, Xinyu and He, Zuyuan and Yan, Lianshan},
	doi = {10.1038/s41377-022-01067-1},
	journal = {Light: Science \& Applications},
	month = {01},
	title = {Integrated sensing and communication in an optical fibre},
	volume = {12},
	year = {2023}}

@inbook{SenCommu_integrated_Review,
	author = {Sabri, Naseer and Aljunid, S. and Salim, Muhammed and Fouad, Sarah},
	doi = {10.1007/978-981-287-128-2_19},
	isbn = {978-981-287-127-5},
	journal = {Springer Series in Materials Science},
	month = {12},
	pages = {299-311},
	title = {Fiber Optic Sensors: Short Review and Applications},
	volume = {204},
	year = {2015}}

@article{SenCommu_integrated_Distributed,
	author = {Yan, Yaxi and Zheng, Hua and Zhao, Zhiyong and Guo, Changjian and Wu, Xiong and Hu, Junhui and Lau, Alan Pak Tao and Lu, Chao},
	doi = {10.1109/JLT.2021.3057670},
	journal = {Journal of Lightwave Technology},
	number = {12},
	pages = {3654-3670},
	title = {Distributed Optical Fiber Sensing Assisted by Optical Communication Techniques},
	volume = {39},
	year = {2021}}

@article{QKD_compen_Simple,
	author = {Agnesi, Costantino and Avesani, Marco and Calderaro, Luca and Stanco, Andrea and Foletto, Giulio and Zahidy, Mujtaba and Scriminich, Alessia and Vedovato, Francesco and Vallone, Giuseppe and Villoresi, Paolo},
	doi = {10.1364/OPTICA.381013},
	journal = {Optica},
	month = {04},
	pages = {284-290},
	title = {Simple quantum key distribution with qubit-based synchronization and a self-compensating polarization encoder},
	volume = {7},
	year = {2020}}

@article{QKD_compen_Network,
	author = {Perani{\'c}, Matej and Clark, Marcus and Wang, Rui and Bahrani, Sima and Alia, Obada and Wengerowsky, S{\"o}ren and Radman, Anton and Lon{\v c}ari{\'c}, Martin and Stip{\v c}evi{\'c}, Mario and Rarity, John and Nejabati, Reza and Joshi, Siddarth},
	doi = {10.1140/epjqt/s40507-023-00187-w},
	journal = {EPJ Quantum Technology},
	month = {08},
	title = {A study of polarization compensation for quantum networks},
	volume = {10},
	year = {2023}}

@article{QKD_compen_PC,
	article-number = {463},
	author = {Xing, Zhuangzhuang and Li, Xingqiao and Ruan, Xinchao and Luo, Yong and Zhang, Hang},
	doi = {10.3390/photonics9070463},
	issn = {2304-6732},
	journal = {Photonics},
	number = {7},
	title = {Phase Compensation for Continuous Variable Quantum Key Distribution Based on Convolutional Neural Network},
	url = {https://www.mdpi.com/2304-6732/9/7/463},
	volume = {9},
	year = {2022}}

@article{QKD_compen_PCF,
	author = {Shiyu Wang and Peng Huang and Miaomiao Liu and Tao Wang and Ping Wang and Guihua Zeng},
	doi = {10.1364/OE.387402},
	journal = {Opt. Express},
	month = {Apr},
	number = {8},
	pages = {10737--10745},
	publisher = {Optica Publishing Group},
	title = {Phase compensation for free-space continuous-variable quantum key distribution},
	url = {https://opg.optica.org/oe/abstract.cfm?URI=oe-28-8-10737},
	volume = {28},
	year = {2020}}

@article{QKD_robust_Practical,
	author = {Sasaki, Toshihiko and Yamamoto, Yoshihisa and Koashi, Masato},
	doi = {10.1038/nature13303},
	journal = {Nature},
	month = {05},
	pages = {475-8},
	title = {Practical quantum key distribution protocol without monitoring signal disturbance},
	volume = {509},
	year = {2014}}

@article{QKD_robust_90km,
	author = {Shimizu, Kaoru and Honjo, Toshimori and Fujiwara, Mikio and Ito, Toshiyuki and Tamaki, Kiyoshi and Miki, Shigehito and Yamashita, Taro and Terai, Hirotaka and Wang, Zhen and Sasaki, Masahide},
	doi = {10.1109/JLT.2013.2291391},
	journal = {Journal of Lightwave Technology},
	number = {1},
	pages = {141-151},
	title = {Performance of Long-Distance Quantum Key Distribution Over 90-km Optical Links Installed in a Field Environment of Tokyo Metropolitan Area},
	volume = {32},
	year = {2014}}

@article{QKD_robust_CV,
	author = {Zhao, Huanxi and Wang, Tao and Xu, Yuehan and Li, Lang and Tan, Zicong and Tan, Piao and Huang, Peng and Zeng, Guihua},
	doi = {10.1364/OE.510392},
	journal = {Optics Express},
	month = {02},
	pages = {7783-7799},
	title = {Continuous-variable quantum key distribution robust against environmental disturbances},
	volume = {32},
	year = {2024}}

@article{QKD_robust_Waters,
	author = {Ata, Yal{\c c}ın and Kiasaleh, Kamran},
	doi = {10.1109/JOE.2025.3551076},
	journal = {IEEE Journal of Oceanic Engineering},
	number = {3},
	pages = {2381-2393},
	title = {Impact of Natural Turbulent Waters on Quantum Key Distribution: Temperature and Salinity Considerations},
	volume = {50},
	year = {2025}}

@article{QKD_compen_gigahertz,
	author = {Shuang Wang and Wei Chen and Zhen-Qiang Yin and De-Yong He and Cong Hui and Peng-Lei Hao and Guan-Jie Fan-Yuan and Chao Wang and Li-Jun Zhang and Jie Kuang and Shu-Feng Liu and Zheng Zhou and Yong-Gang Wang and Guang-Can Guo and Zheng-Fu Han},
	doi = {10.1364/OL.43.002030},
	journal = {Opt. Lett.},
	month = {May},
	number = {9},
	pages = {2030--2033},
	publisher = {Optica Publishing Group},
	title = {Practical gigahertz quantum key distribution robust against channel disturbance},
	url = {https://opg.optica.org/ol/abstract.cfm?URI=ol-43-9-2030},
	volume = {43},
	year = {2018}}

@article{Sagnac_appl_PM,
	author = {Luo, Wenbin and Li, Yang and Li, Yuhuai and Tao, Xueying and Han, Liying and Cai, Wenqi and Yin, Juan and Ren, Jigang and Liao, Shengkai and Peng, Chengzhi},
	doi = {10.1109/JPHOT.2022.3209532},
	journal = {IEEE Photonics Journal},
	number = {5},
	pages = {1-6},
	title = {Intrinsically Stable 2-GHz Polarization Modulation for Satellite-Based Quantum Key Distribution},
	volume = {14},
	year = {2022}}

@article{Sagnac_appl_CVmodulator,
	author = {Huanxi Zhao and Huasheng Li and Yuehan Xu and Peng Huang and Tao Wang and Guihua Zeng},
	doi = {10.1364/OL.458443},
	journal = {Opt. Lett.},
	month = {Jun},
	number = {12},
	pages = {2939--2942},
	publisher = {Optica Publishing Group},
	title = {Simple continuous-variable quantum key distribution scheme using a Sagnac-based Gaussian modulator},
	url = {https://opg.optica.org/ol/abstract.cfm?URI=ol-47-12-2939},
	volume = {47},
	year = {2022}}

@article{Sagnac_appl_CVfree,
	author = {Xue-Tao Zheng and Qi-Fa Zhang and Jing-yu Han and Jie Ling and Guang-can Guo and Zheng-Fu Han},
	doi = {10.1364/OL.502897},
	journal = {Opt. Lett.},
	month = {Sep},
	number = {18},
	pages = {4837--4840},
	publisher = {Optica Publishing Group},
	title = {Experimental realization of free-space continuous-variable quantum key distribution based on fiber Sagnac interferometer},
	url = {https://opg.optica.org/ol/abstract.cfm?URI=ol-48-18-4837},
	volume = {48},
	year = {2023}}

@article{Sagnac_appl_source,
	author = {Singh, Sandeep and Kumar, Vimlesh and Ghosh, Anirban and Forbes, Andrew and Samanta, Goutam K},
	doi = {10.1002/qute.202200121},
	journal = {Advanced Quantum Technologies},
	month = {12},
	title = {A Tolerance‐Enhanced Spontaneous Parametric Downconversion Source of Bright Entangled Photons},
	volume = {6},
	year = {2022}}

@article{ZF_ref,
	author = {Hoffman, P.R. and Kuzyk, M.G.},
	doi = {10.1109/JLT.2004.824455},
	journal = {Journal of Lightwave Technology},
	number = {2},
	pages = {494-498},
	title = {Position determination of an acoustic burst along a Sagnac interferometer},
	volume = {22},
	year = {2004}}

@article{Sagnc_QKD_TD,
	author = {Zhou, Chunyuan and Zeng, Heping},
	doi = {10.1063/1.1541102},
	eprint = {https://pubs.aip.org/aip/apl/article-pdf/82/5/832/18575632/832_1_online.pdf},
	issn = {0003-6951},
	journal = {Applied Physics Letters},
	month = {02},
	number = {5},
	pages = {832-834},
	title = {Time-division single-photon Sagnac interferometer for quantum key distribution},
	url = {https://doi.org/10.1063/1.1541102},
	volume = {82},
	year = {2003}}

@article{Sagnc_QKD_Telecom,
	author = {Jan Bogdanski and Johan Ahrens and Mohamed Bourennane},
	doi = {https://doi.org/10.1016/j.optcom.2008.12.023},
	issn = {0030-4018},
	journal = {Optics Communications},
	number = {6},
	pages = {1231-1236},
	title = {Sagnac quantum key distribution over telecom fiber networks},
	url = {https://www.sciencedirect.com/science/article/pii/S0030401808012704},
	volume = {282},
	year = {2009}}

@article{Sagnc_QKD_LKG,
	author = {Xu, Feihu and Qi, Bing and Lo, Hoi-Kwong},
	doi = {10.1088/1367-2630/12/11/113026},
	journal = {New Journal of Physics},
	month = {nov},
	number = {11},
	pages = {113026},
	title = {Experimental demonstration of phase-remapping attack in a practical quantum key distribution system},
	url = {https://doi.org/10.1088/1367-2630/12/11/113026},
	volume = {12},
	year = {2010}}

@inproceedings{Sagnc_QKD_TF,
	author = {Xiaoqing Zhong and Wenyuan Wang and Li Qian and Hoi-Kwong Lo},
	booktitle = {Conference on Lasers and Electro-Optics},
	doi = {10.1364/CLEO_QELS.2020.FF3C.1},
	journal = {Conference on Lasers and Electro-Optics},
	pages = {FF3C.1},
	publisher = {Optica Publishing Group},
	title = {Proof-of-principle experimental demonstration of twin-field quantum key distribution over asymmetric channels},
	url = {https://opg.optica.org/abstract.cfm?URI=CLEO_QELS-2020-FF3C.1},
	year = {2020}}

@article{WM_our_work,
	author = {Zhao, Wei-Qian and Su, Zi-Fu and Yu, Ya-Fei and Wang, Jin-Dong},
	doi = {10.1109/JLT.2025.3565356},
	journal = {Journal of Lightwave Technology},
	number = {14},
	pages = {6948-6956},
	title = {Long-Distance High-Precision and High-Sensitivity Time Delay Variation Sensing Based on Fiber-Optic Weak Measurements},
	volume = {43},
	year = {2025}}

@article{QKD_NW_Ring_distribution,
	author = {Yuan Ren and Xutong Wang and Yinghui Lv and Davide Bacco and Jietai Jing},
	journal = {Laser \& Photonics Reviews},
	title = {Distribution of Multiplexed Continuous‐Variable Entanglement for Quantum Networks},
	url = {https://api.semanticscholar.org/CorpusID:251878830},
	volume = {16},
	year = {2022}}

@article{QKD_NW_Ring_study,
	author = {{Perani{\'c}, Matej} and {Clark, Marcus} and {Wang, Rui} and {Bahrani, Sima} and {Alia, Obada} and {Wengerowsky, S{\"o}ren} and {Radman, Anton} and {Lon{\v c}ari{\'c}, Martin} and {Stip{\v c}evi{\'c}, Mario} and {Rarity, John} and {Nejabati, Reza} and {Joshi, Siddarth Koduru}},
	doi = {10.1140/epjqt/s40507-023-00187-w},
	journal = {EPJ Quantum Technol.},
	number = 1,
	pages = {30},
	title = {A study of polarization compensation for quantum networks},
	url = {https://doi.org/10.1140/epjqt/s40507-023-00187-w},
	volume = 10,
	year = 2023}

@inproceedings{QKD_NW_Ring_muti,
	author = {Bing C. Wang and Patrick Kumavor and Susanne F. Yelin and Alan C. Beal},
	booktitle = {Active and Passive Optical Components for WDM Communications V},
	doi = {10.1117/12.632251},
	editor = {Achyut K. Dutta and Yasutake Ohishi and Niloy K. Dutta and Jesper Moerk},
	organization = {International Society for Optics and Photonics},
	pages = {601416},
	publisher = {SPIE},
	title = {{Multi-user quantum cryptography}},
	url = {https://doi.org/10.1117/12.632251},
	volume = {6014},
	year = {2005}}

@article{QKD_NW_Star_LYM,
	author = {Liu, Shuaishuai and Lu, Zhenguo and Wang, Pu and Tian, Yan and Wang, Xuyang and Li, Yongmin},
	doi = {10.1038/s41534-023-00763-z},
	journal = {npj Quantum Information},
	month = {09},
	title = {Experimental demonstration of multiparty quantum secret sharing and conference key agreement},
	volume = {9},
	year = {2023}}

@article{QKD_NW_Star_S_G,
	author = {Yu-Ao Chen and Qiang Zhang and Teng-Yun Chen and Wenqi Cai and Shengkai Liao and Jun Zhang and Kai Chen and Juan Yin and Ji-Gang Ren and Zhu Chen and Sheng Long Han and Qingyang Yu and Ken Liang and Fei Zhou and Xiao Yuan and Mei-sheng Zhao and Tianyin Wang and Xiao Jiang and Liang Zhang and Weiyue Liu and Yang Li and Qi Shen and Yuan Cao and Chaoyang Lu and Rong Shu and Jian-Yu Wang and Li Li and Nai-Le Liu and Feihu Xu and Xiang-Bin Wang and Cheng-Zhi Peng and Jian-Wei Pan},
	journal = {Nature},
	pages = {214 - 219},
	title = {An integrated space-to-ground quantum communication network over 4,600 kilometres},
	url = {https://api.semanticscholar.org/CorpusID:230812317},
	volume = {589},
	year = {2021}}

@article{QKD_NW_Star_access,
	author = {Fr{\"o}hlich, Bernd and Dynes, James and Lucamarini, Marco and Sharpe, Andrew and Yuan, Zhiliang and Shields, Andrew},
	doi = {10.1038/nature12493},
	journal = {Nature},
	month = {09},
	pages = {69-72},
	title = {A quantum access network},
	volume = {501},
	year = {2013}}

@article{QKD_NW_Star_ZGH,
	author = {Xu, Yuehan and Wang, Tao and Zhao, Huanxi and Huang, Peng and Zeng, Guihua},
	doi = {10.1364/PRJ.492448},
	journal = {Photonics Research},
	month = {06},
	title = {Round-trip multi-band quantum access network},
	volume = {11},
	year = {2023}}

@article{QKD_NW_P_P_Swiss,
	author = {Stucki, D and Legr{\'e}, M and Buntschu, F and Clausen, B and Felber, N and Gisin, N and Henzen, L and Junod, P and Litzistorf, G and Monbaron, P and Monat, L and Page, J-B and Perroud, D and Ribordy, G and Rochas, A and Robyr, S and Tavares, J and Thew, R and Trinkler, P and Ventura, S and Voirol, R and Walenta, N and Zbinden, H},
	doi = {10.1088/1367-2630/13/12/123001},
	journal = {New Journal of Physics},
	month = {dec},
	number = {12},
	pages = {123001},
	publisher = {IOP Publishing},
	title = {Long-term performance of the SwissQuantum quantum key distribution network in a field environment},
	url = {https://doi.org/10.1088/1367-2630/13/12/123001},
	volume = {13},
	year = {2011}}

@article{QKD_NW_P_P_Tokyo,
	author = {M. Sasaki and M. Fujiwara and H. Ishizuka and W. Klaus and K. Wakui and M. Takeoka and S. Miki and T. Yamashita and Z. Wang and A. Tanaka and K. Yoshino and Y. Nambu and S. Takahashi and A. Tajima and A. Tomita and T. Domeki and T. Hasegawa and Y. Sakai and H. Kobayashi and T. Asai and K. Shimizu and T. Tokura and T. Tsurumaru and M. Matsui and T. Honjo and K. Tamaki and H. Takesue and Y. Tokura and J. F. Dynes and A. R. Dixon and A. W. Sharpe and Z. L. Yuan and A. J. Shields and S. Uchikoga and M. Legr\'{e} and S. Robyr and P. Trinkler and L. Monat and J.-B. Page and G. Ribordy and A. Poppe and A. Allacher and O. Maurhart and T. L\"{a}nger and M. Peev and A. Zeilinger},
	doi = {10.1364/OE.19.010387},
	journal = {Opt. Express},
	month = {May},
	number = {11},
	pages = {10387--10409},
	publisher = {Optica Publishing Group},
	title = {Field test of quantum key distribution in the Tokyo QKD Network},
	url = {https://opg.optica.org/oe/abstract.cfm?URI=oe-19-11-10387},
	volume = {19},
	year = {2011}}

@article{QKD_NW_P_P_Field,
	author = {Duan Huang and Peng Huang and Huasheng Li and Tao Wang and Yingming Zhou and Guihua Zeng},
	doi = {10.1364/OL.41.003511},
	journal = {Opt. Lett.},
	month = {Aug},
	number = {15},
	pages = {3511--3514},
	publisher = {Optica Publishing Group},
	title = {Field demonstration of a continuous-variable quantum key distribution network},
	url = {https://opg.optica.org/ol/abstract.cfm?URI=ol-41-15-3511},
	volume = {41},
	year = {2016}}

@article{QKD_NW_P_P_upstream,
	author = {Xiangyu Wang and Ziyang Chen and Zhenghua Li and Dengke Qi and Song Yu and Hong Guo},
	doi = {10.1364/OL.487582},
	journal = {Opt. Lett.},
	month = {Jun},
	number = {12},
	pages = {3327--3330},
	publisher = {Optica Publishing Group},
	title = {Experimental upstream transmission of continuous variable quantum key distribution access network},
	url = {https://opg.optica.org/ol/abstract.cfm?URI=ol-48-12-3327},
	volume = {48},
	year = {2023}}

@article{Qua_Cla_advance_intergrated,
	author = {Orieux, Adeline and Diamanti, Eleni},
	doi = {10.1088/2040-8978/18/8/083002},
	journal = {Journal of Optics},
	month = {jul},
	number = {8},
	pages = {083002},
	publisher = {IOP Publishing},
	title = {Recent advances on integrated quantum communications},
	url = {https://doi.org/10.1088/2040-8978/18/8/083002},
	volume = {18},
	year = {2016}}

@article{Qua_Cla_Novel,
	author = {Fabrizio Granelli and Riccardo Bassoli and Janis N{\"o}tzel and Frank H. P. Fitzek and Holger Boche and Nelson L. S. da Fonseca},
	journal = {Wireless Communications and Mobile Computing},
	title = {A Novel Architecture for Future Classical-Quantum Communication Networks},
	url = {https://api.semanticscholar.org/CorpusID:248446487},
	year = {2022}}

@article{Qua_Cla_Hybrid,
	author = {Joseph M. Lukens and Nicholas A. Peters and Bing Qi},
	doi = {https://doi.org/10.1016/j.pquantelec.2025.100586},
	issn = {0079-6727},
	journal = {Progress in Quantum Electronics},
	pages = {100586},
	title = {Hybrid classical-quantum communication networks},
	url = {https://www.sciencedirect.com/science/article/pii/S0079672725000345},
	volume = {103},
	year = {2025}}

@article{Qua_Cla_interplay,
	author = {V{\'a}zquez-Castro, Angeles and Han, Zhu},
	doi = {10.1109/MCOM.009.2300856},
	journal = {IEEE Communications Magazine},
	number = {10},
	pages = {54-60},
	title = {Interplay of Classical-Quantum Resources in Space-Terrestrial Integrated Networks},
	volume = {62},
	year = {2024}}

@article{Qua_Cla_enhanced,
	author = {Zhang, Wen-Hao and Liu, Xiao and Yin, Peng and Peng, Xing-Xiang and Li, Gong-Chu and Xu, Xiao-Ye and Yu, Shang and Hou, Zhi-Bo and Han, Yong-Jian and Xu, Jin-Shi and Zhou, Zong-Quan and Chen, Geng and Li, Chuan-Feng and Guo, Guang-Can},
	doi = {10.1038/s41534-020-00328-4},
	journal = {npj Quantum Information},
	month = {12},
	title = {Classical communication enhanced quantum state verification},
	volume = {6},
	year = {2020}}

@unknown{Qua_Cla_Wireless,
	author = {Popovski, Petar and Stefanovi{\'c}, {\v C}edomir and Soret, Beatriz and Leyva-Mayorga, Israel and Pandey, Shashi and Christensen, Ren{\'e} and S{\o}ndergaard, Jakob and Jensen, Kristian and Pedersen, Thomas and Cacciapuoti, Angela Sara and Hanzo, Lajos},
	doi = {10.48550/arXiv.2509.14731},
	month = {09},
	title = {1Q: First-Generation Wireless Systems Integrating Classical and Quantum Communication},
	year = {2025}}

@inproceedings{Qua_Cla_network,
	author = {Sharan, Ayush Kumar and Philip, Alwin and Sharma, Purushottam and Shukla, Vinod Kumar},
	booktitle = {2024 11th International Conference on Reliability, Infocom Technologies and Optimization (Trends and Future Directions) (ICRITO)},
	doi = {10.1109/ICRITO61523.2024.10522244},
	pages = {1-8},
	title = {Quantum Networking: Creating the Future Landscape of Refined Classical Communication},
	year = {2024}}

@article{Qua_Cla_multicore_fibers,
	author = {Wu, Qi and Ribezzo, Domenico and Sciullo, Giammarco and Cocchi, Sebastiano and Shaji, Divya and Alves Zischler, Lucas and Luis, Ruben and Serena, Paolo and Lasagni, Chiara and Bononi, A. and Hayashi, Tetsuya and Gagliano, Alessandro and Martelli, Paolo and Gatto, Alberto and Parolari, Paola and Boffi, Pierpaolo and Bacco, Davide and Zavatta, Alessandro and Zhu, Yixiao and Antonelli, Cristian},
	doi = {10.1038/s41377-025-01982-z},
	journal = {Light, science \& applications},
	month = {08},
	pages = {274},
	title = {Integration of quantum key distribution and high-throughput classical communications in field-deployed multi-core fibers},
	volume = {14},
	year = {2025}}

@article{QKD_Conjugate_Conding,
	address = {New York, NY, USA},
	author = {Wiesner, Stephen},
	doi = {10.1145/1008908.1008920},
	issn = {0163-5700},
	issue_date = {Winter-Spring 1983},
	journal = {SIGACT News},
	month = jan,
	number = {1},
	numpages = {11},
	pages = {78--88},
	publisher = {Association for Computing Machinery},
	title = {Conjugate coding},
	url = {https://doi.org/10.1145/1008908.1008920},
	volume = {15},
	year = {1983}}

@article{QKD_BB,
	author = {Charles H. Bennett and Gilles Brassard},
	doi = {https://doi.org/10.1016/j.tcs.2014.05.025},
	issn = {0304-3975},
	journal = {Theoretical Computer Science},
	note = {Theoretical Aspects of Quantum Cryptography -- celebrating 30 years of BB84},
	pages = {7-11},
	title = {Quantum cryptography: Public key distribution and coin tossing},
	url = {https://www.sciencedirect.com/science/article/pii/S0304397514004241},
	volume = {560},
	year = {2014}}

@article{QKD_Advance,
	author = {S. Pirandola and U. L. Andersen and L. Banchi and M. Berta and D. Bunandar and R. Colbeck and D. Englund and T. Gehring and C. Lupo and C. Ottaviani and J. L. Pereira and M. Razavi and J. Shamsul Shaari and M. Tomamichel and V. C. Usenko and G. Vallone and P. Villoresi and P. Wallden},
	doi = {10.1364/AOP.361502},
	journal = {Adv. Opt. Photon.},
	month = {Dec},
	number = {4},
	pages = {1012--1236},
	publisher = {Optica Publishing Group},
	title = {Advances in quantum cryptography},
	url = {https://opg.optica.org/aop/abstract.cfm?URI=aop-12-4-1012},
	volume = {12},
	year = {2020}}

@article{QKD_PCR,
	author = {Portmann, Christopher and Renner, Renato},
	doi = {10.1103/RevModPhys.94.025008},
	issue = {2},
	journal = {Rev. Mod. Phys.},
	month = {Jun},
	numpages = {56},
	pages = {025008},
	publisher = {American Physical Society},
	title = {Security in quantum cryptography},
	url = {https://link.aps.org/doi/10.1103/RevModPhys.94.025008},
	volume = {94},
	year = {2022}}

@article{QKD_LKG,
	author = {Xu, Feihu and Ma, Xiongfeng and Zhang, Qiang and Lo, Hoi-Kwong and Pan, Jian-Wei},
	doi = {10.1103/RevModPhys.92.025002},
	issue = {2},
	journal = {Rev. Mod. Phys.},
	month = {May},
	numpages = {60},
	pages = {025002},
	publisher = {American Physical Society},
	title = {Secure quantum key distribution with realistic devices},
	url = {https://link.aps.org/doi/10.1103/RevModPhys.92.025002},
	volume = {92},
	year = {2020}}

@article{Integrated_Polan,
	author = {Marek {\.Z}yczkowski and Marcin Kowalski},
	journal = {Acta Physica Polonica A},
	pages = {606-609},
	title = {A Quantum Key as the Fiber Optic Security Sensor},
	url = {https://api.semanticscholar.org/CorpusID:110880237},
	volume = {124},
	year = {2013}}

@article{Integrated_Shapiro,
	author = {Zhuang, Quntao and Zhang, Zheshen and Shapiro, Jeffrey H.},
	doi = {10.1103/PhysRevA.97.032329},
	issue = {3},
	journal = {Phys. Rev. A},
	month = {Mar},
	numpages = {6},
	pages = {032329},
	publisher = {American Physical Society},
	title = {Distributed quantum sensing using continuous-variable multipartite entanglement},
	url = {https://link.aps.org/doi/10.1103/PhysRevA.97.032329},
	volume = {97},
	year = {2018}}

@article{Integrated_GuoGC,
	author = {Yin, Peng and Takeuchi, Yuki and Zhang, Wen-Hao and Yin, Zhen-Qiang and Matsuzaki, Yuichiro and Peng, Xing-Xiang and Xu, Xiao-Ye and Xu, Jin-Shi and Tang, Jian-Shun and Zhou, Zong-Quan and Chen, Geng and Li, Chuan-Feng and Guo, Guang-Can},
	doi = {10.1103/PhysRevApplied.14.014065},
	issue = {1},
	journal = {Phys. Rev. Appl.},
	month = {Jul},
	numpages = {9},
	pages = {014065},
	publisher = {American Physical Society},
	title = {Experimental Demonstration of Secure Quantum Remote Sensing},
	url = {https://link.aps.org/doi/10.1103/PhysRevApplied.14.014065},
	volume = {14},
	year = {2020}}

@article{Integrated_PanJW,
	author = {Chen, Jiu-Peng and Zhang, Chi and Liu, Yang and Jiang, Cong and Zhao, Dong-Feng and Zhang, Wei-Jun and Chen, Fa-Xi and Li, Hao and You, Li-Xing and Wang, Zhen and Chen, Yang and Wang, Xiang-Bin and Zhang, Qiang and Pan, Jian-Wei},
	doi = {10.1103/PhysRevLett.128.180502},
	issue = {18},
	journal = {Phys. Rev. Lett.},
	month = {May},
	numpages = {6},
	pages = {180502},
	publisher = {American Physical Society},
	title = {Quantum Key Distribution over 658 km Fiber with Distributed Vibration Sensing},
	url = {https://link.aps.org/doi/10.1103/PhysRevLett.128.180502},
	volume = {128},
	year = {2022}}

@article{Integrated_LiYM,
	author = {Shuaishuai Liu and Yan Tian and Yu Zhang and Zhenguo Lu and Xuyang Wang and Yongmin Li},
	journal = {Optica},
	title = {Integrated quantum communication network and vibration sensing in optical fibers},
	url = {https://api.semanticscholar.org/CorpusID:268793386},
	year = {2024}}

@article{Integrated_ZhengGH,
	author = {Yuehan Xu and Tao Wang and Peng Huang and Guihua Zeng},
	doi = {10.34133/research.0416},
	eprint = {https://spj.science.org/doi/pdf/10.34133/research.0416},
	journal = {Research},
	pages = {0416},
	title = {Integrated Distributed Sensing and Quantum Communication Networks},
	url = {https://spj.science.org/doi/abs/10.34133/research.0416},
	volume = {7},
	year = {2024}}

@article{Integrated_IoT,
	author = {Liao, Haijun and Li, Ziming and Ci, Haoyu and Zhang, Sunxuan and Yao, Zijia and Zhou, Zhenyu and Mumtaz, Shahid},
	doi = {10.1109/TSG.2025.3599386},
	journal = {IEEE Transactions on Smart Grid},
	pages = {1-1},
	title = {Integrated Sensing, Transmission and Control for QKD-IoT Empowered Power-Communication Coupling Smart Grid},
	year = {2025}}

\end{document}